\documentclass[11pt]{article}
\usepackage[htt]{hyphenat}
\usepackage[T1]{fontenc}
\usepackage[utf8]{inputenc}
\usepackage{authblk}
\usepackage{setspace}
\usepackage{geometry}
\usepackage{enumitem}
\usepackage{url}
\usepackage[numbers,super,sort&compress]{natbib}
\usepackage{graphicx}
\usepackage{booktabs}
\usepackage{makecell}
\usepackage{siunitx}
\usepackage{enumitem}
\usepackage{url}
\usepackage[table]{xcolor}
\usepackage{subcaption, tabularx}
\usepackage{listings}
\usepackage{xcolor}

\usepackage[section]{placeins}
\title{Unifying points of interest taxonomies: mapping OpenStreetMap tags to the Foursquare category system}

\author[1]{Lilou Soulas}
\author[2,3]{Lorenzo Lucchini}
\author[2]{Maurizio Napolitano}
\author[2]{Sebastiano Bontorin}
\author[2]{Simone Centellegher}
\author[2]{Bruno Lepri}
\author[2]{Riccardo Gallotti}
\author[2,*]{Eleonora Andreotti}
\affil[1]{ENSAE Paris, École Nationale de la Statistique et de l'Administration Économique, Institut Polytechnique de Paris, France}
\affil[2]{Fondazione Bruno Kessler (FBK), Povo (TN), Italy}
\affil[3]{Bocconi University, DONDENA and BIDSA research centers, Milan, Italy.}
\affil[*]{Corresponding author (e-mail: eandreotti@fbk.eu).}
\date{} 
\begin{document}
\maketitle

\begin{abstract}
The heterogeneity of Point of Interest (POI) taxonomies is a persistent challenge for the integration of urban datasets and the development of location-based services. OpenStreetMap (OSM) adopts a flexible, community-driven tagging system, while Foursquare (FS) relies on a curated hierarchical structure. Here we present an openly available benchmark and mapping framework that aligns OSM tags with the FS taxonomy. This resource integrates the richness of community-driven OSM data with the hierarchical structure of FS, enabling reproducible and interoperable urban analytics. The dataset is complemented by an evaluation of embedding and LLM-based alignment strategies and a pipeline that supports scalable updates as OSM evolves. Together, these elements provide both a robust reference resource and a practical tool for the community. Our approach is structured around three components: the construction of a manually curated benchmark as a gold standard, the evaluation of pretrained text embedding models for semantic alignment between OSM tags and FS categories, and an LLM-based refinement stage that enhances robustness and adaptability. The proposed methodology provides a scalable and reproducible solution for taxonomy unification, with direct applications to urban analytics, mobility studies, and smart city services.
\end{abstract}


\section*{Background \& Summary}
The integration of heterogeneous urban datasets is crucial for a more comprehensive understanding of socio-urban systems \cite{lucchini2021living,moro2021mobility,fraser2024great,centellegher2025job,bontorin2025mixing,Li2023StreetGreenery}, and the development of data-driven applications in mobility, urban planning, and smart city services \cite{andreotti2025city}. 
A wide range of datasets currently provides geolocated information on urban functions and Points of Interest (POIs). These sources differ substantially in their data collection methods, licensing regimes, and reliability. Fully proprietary platforms such as {Google Places} \cite{googleplaces} and {SafeGraph} \cite{safegraph} offer extensive global coverage and rich metadata, including business hours, popularity metrics, and detailed categorical hierarchies, but operate under restrictive licenses that limit large-scale access, reuse, and reproducibility. Their APIs are closed and their update frequencies are not publicly disclosed, which may limit transparency and independent validation. Conversely, open or collaborative initiatives such as {OpenStreetMap (OSM)}, the {Foursquare Open Source Places} dataset, and the {Overture Maps Foundation (OMF)} provide accessible and reusable data under open licenses. OSM, distributed under the {Open Database License (ODbL)}, is fully community-driven and continuously updated, though its quality can vary across regions depending on contributor density \cite{haklay2008osm}. Foursquare’s {Open Source Places} dataset, released under the {Apache 2.0} license, offers global coverage and a curated taxonomy of POIs, albeit with fewer available attributes than its commercial counterpart \cite{foursquare_open}, \cite{foursquare}. The {Overture Maps Foundation}, an open consortium led by Meta, Microsoft, Amazon AWS, and TomTom, aggregates data from multiple sources, including OSM and corporate contributions, under a {CDLA Permissive 2.0} license, harmonizing schemas and identifiers while still facing challenges of completeness and duplication in its early stages of adoption \cite{overture}. Within this heterogeneous landscape, we focus on {Foursquare (FS)} and {OpenStreetMap (OSM)} because they embody two complementary paradigms of spatial data production, curated versus community-generated, that are both accessible for research and widely adopted in urban and computational social science. OSM captures the diversity and spontaneity of bottom-up mapping practices, while Foursquare provides a top-down, hierarchically organized view of urban activities. Together, these datasets enable comparative analyses of how different data-generation logics and licensing frameworks shape the semantic and spatial representation of urban functions. According to a search in the OpenAlex database of works (title + abstract) from 2001 onwards, {3 693} scientific outputs (papers, books, datasets) mention ``Foursquare'' and {5 109} mention ``OpenStreetMap''. The number of OSM-related works shows a gradual increase over time, surpassing 600 publications in 2023, whereas Foursquare-related outputs peaked between 2013 and 2017 before showing a slight decline. When restricting the search to the expressions “Foursquare categories” and “OpenStreetMap tags”, the results narrow to {152} and {308} works, respectively \cite{priem2022openalex, openalex2025}. Although both resources are extensively used in research and practice, their heterogeneity hampers interoperability and limits the scalability of location-based services.\\
This heterogeneity does not only make integration difficult, but also obscures important semantic relationships between datasets. In many cases, OSM and FS categories describe the same type of object under different labels, leading to redundancies that prevent unified analysis. At the same time, some categories that are well defined in one taxonomy are absent or too coarse in the other, revealing potential gaps that limit the richness of urban datasets. Addressing these overlaps and omissions is therefore essential not only for harmonizing existing data, but also for informing the design of more comprehensive and granular taxonomies.\\
Existing approaches to taxonomy alignment have largely relied on manual curation, lexical similarity measures, or ontology-based mappings (see \citet{zhu2012survey} for a survey of these approaches). While effective in small-scale settings, these methods face three main limitations: they are not scalable when applied to the tens of thousands of OSM tags; they are brittle to the dynamic evolution of OSM, where new tags and categories are continuously introduced; and they often fail to capture the semantic nuance required for robust alignment with structured systems such as FS.\\
Some recent studies have attempted to automatically construct hierarchical taxonomies directly from OSM data \cite{shbita2024osm}. While these approaches help impose structure on the OSM tagging system, they face several intrinsic limitations: they tend to discard rare but semantically relevant tags, often generate redundant or inconsistent parent-child relations, and remain highly sensitive to regional idiosyncrasies (for instance, when semantically equivalent features are tagged differently across countries, such as \texttt{amenity=pub} versus \texttt{amenity=bar}, or \texttt{shop=chemist} versus \texttt{shop=pharmacy}, reflecting cultural and linguistic variations in mapping conventions). Moreover, the resulting hierarchies are validated only qualitatively and require human intervention to ensure coherence. Most importantly, they remain confined to OSM itself and do not address the broader challenge of establishing interoperability across heterogeneous POI taxonomies. In particular, they neither provide mappings to curated systems such as FS nor identify missing or overlapping categories across datasets, which are crucial requirements for unified urban analytics.\\
Other works have explicitly compared OSM with FS. For instance, Zhang and Pfoser \cite{zhang2019osmfs} investigated the feasibility of using OSM and FS POIs to model urban change, analyzing coverage differences and spatial distributions. Novack et al. \cite{novack2018graph} proposed a graph-based approach to match POIs between collaborative geo-datasets, including OSM and FS, primarily relying on name similarity and spatial proximity. While these studies highlight important complementarities between the two sources, they do not provide a systematic or scalable taxonomy-level alignment, leaving unresolved the problem of unifying heterogeneous POI categorizations.\\
The advent of Large Language Models (LLMs) has opened new opportunities for semantic matching tasks, as these models encode rich contextual information and generalize across domains. Recent studies have applied LLMs to entity matching \cite{11107461}, ontology alignment \cite{10.1145/3587259.3627571, TABOADA2025103254}, and classification problems \cite{LeHaNguyenEtAl2025}, demonstrating improved performance over traditional similarity-based methods. However, to the best of our knowledge, the problem of unifying heterogeneous POI taxonomies using LLMs remains underexplored.\\
In this paper, we introduce a framework that combines pretrained sentence-transformer models with a generative LLM (ChatGPT-4o-mini \cite{openai_gpt4omini}, accessed via the OpenAI API) to automatically align OSM tags with the FS category system\footnote{We also tested ChatGPT-5-mini \cite{openai_gpt5mini}, which yielded comparable results but increased computational cost.}. 
The FS taxonomy provides a curated hierarchical structure that enables analyses at multiple levels of granularity, ranging from broad thematic groups to fine-grained categories (e.g., \texttt{Dining and Drinking} > \texttt{Restaurant} > \texttt{Asian Restaurant} > \texttt{Chinese Restaurant} > \texttt{Dim Sum Restaurant}). Extending this structure to OSM data allows us to integrate the richness of community-driven tagging with the consistency required for reliable multi-scale analyses.

The resulting resource, encompassing both the released dataset and the accompanying framework, offers three main components:
\begin{enumerate}[label=\roman*)]
    \item a manually validated benchmark mapping OSM tags to the FS taxonomy, which provides a reliable gold-standard reference;
    \item an evaluation of embedding- and LLM-based alignment strategies;
    \item a scalable framework that supports updates as OSM evolves, ensuring long-term interoperability.
\end{enumerate}

Together, these elements make the OSM$\rightarrow$FS benchmark not only a robust reference dataset but also a \textit{long-lived resource}, supported by methods and code that enable reproducible evaluation and continuous adaptation.


\section*{Methods}
\label{sec:methodology}

\begin{figure}[h]
    \centering
    \includegraphics[width=1\textwidth]{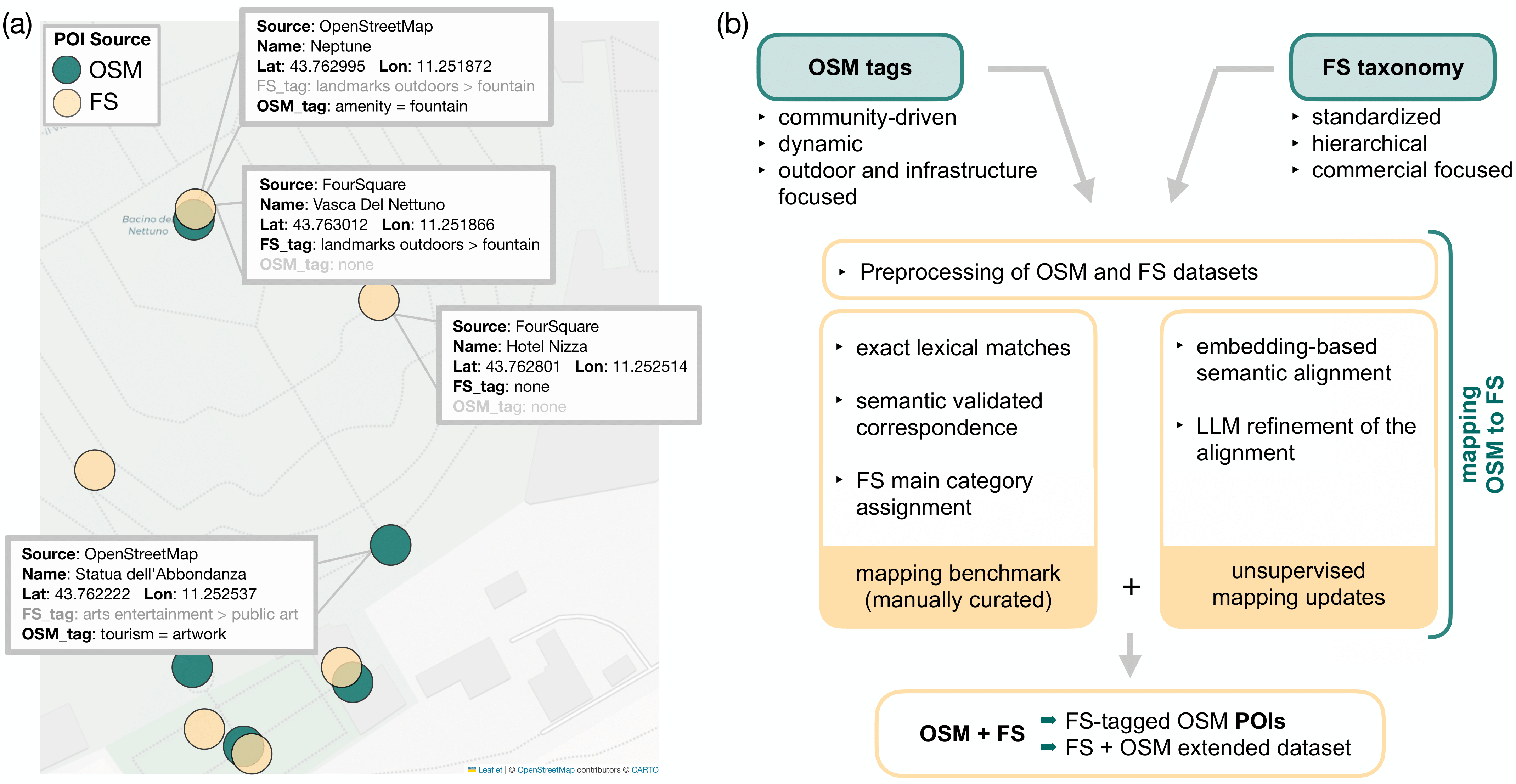}
    \caption{Integration and semantic mapping between OSM and FS data sources.
(a) Examples of corresponding POIs from OSM and FS in the city of Bologna (Italy), showing differences in tagging structure and coverage.
(b) Overview of the OSM-FS mapping pipeline, including exact lexical matching, semantic validation, and embedding-based alignment refined with LLM assistance. The process produces FS-tagged OSM POIs and an extended dataset combining both taxonomies.}
\label{fig:prima}
\end{figure}

Our methodology is structured in three steps (Figure \ref{fig:prima}), reflecting both the construction of a reliable evaluation baseline and the design of an automatic alignment framework. First, we created a manually validated benchmark mapping between OSM and FS categories, which already constitutes a valuable contribution in its own right, as it provides a high-quality reference for subsequent studies. However, since OSM is a dynamic system in which new tags are continuously introduced and existing ones evolve, a purely manual approach is not sufficient to guarantee long-term interoperability. For this reason, we complement the benchmark with two automatic steps. The second step relies on sentence embedding models to capture semantic similarity between categories and to retrieve candidate alignments at scale. While effective, this embedding-based approach alone does not achieve the accuracy required for robust taxonomy unification. Therefore, in the third step we introduce an LLM-based refinement procedure, which leverages the contextual reasoning capabilities of large language models to select the most appropriate mapping among top-ranked candidates. Together, these three steps form a scalable framework that combines the reliability of manual validation with the adaptability of automatic methods.
\subsection*{Data Source}
\label{sec:data}
Before detailing the three methodological steps, we first describe the raw datasets used in our study and the preprocessing procedures required to obtain a consistent and comparable representation of POIs. Specifically, we introduce the Foursquare datasets, comprising the Places and Categories resources, and outline the extraction and cleaning of OpenStreetMap (OSM) tags. These datasets provide the textual inputs that form the foundation for constructing the benchmark mapping and for subsequent alignment methods.
\subsubsection*{Foursquare Dataset}
The FS data were obtained from the official platform and comprise the Places dataset, consisting of approximately 100 Parquet files describing individual POIs, and the Categories dataset, consisting of a single Parquet file containing the category hierarchy.
Each POI file includes about one million records with identifiers, spatiotemporal attributes, and metadata such as name, contact information, and links to social media accounts. Of particular interest is the field \texttt{fsq\_category\_labels}, which encodes the most granular category (or categories) associated with a place (e.g., \texttt{Landmarks and Outdoors > Stable}). 
The FS category file defines a taxonomy of 1,244 categories organized into six hierarchical levels and distributed across 11 top-level categories. After normalization (e.g., removal of lowercase inconsistencies, harmonization of variable names), each POI can be represented as a path from a top-level category (\texttt{Depth\_1}) to its most specific label (\texttt{Depth\_6}). Table~\ref{fig:carte_fs} illustrates the resulting clean hierarchical structure, which provides a consistent and well-defined backbone for urban analytics.

\begin{table*}[htbp]
\centering
\resizebox{\textwidth}{!}{%
\begin{tabular}{l c l l l l l l}
\toprule
\textbf{Tag} & \textbf{Depth} & \textbf{Depth 1} & \textbf{Depth 2} & \textbf{Depth 3} & \textbf{Depth 4} & \textbf{Depth 5} & \textbf{Depth 6} \\
\midrule
kaiseki restaurant     & 5 & Dining and Drinking & Restaurant & Asian restaurant & \makecell[l]{Japanese\\restaurant} & \makecell[l]{Kaiseki\\restaurant} & -- \\
art museum             & 3 & Arts and Entertainment & Museum & Art museum & -- & -- & -- \\
rental car location    & 3 & Travel and Transportation & Transport hub & Rental car location & -- & -- & -- \\
shabu-shabu restaurant & 5 & Dining and Drinking & Restaurant & Asian restaurant & \makecell[l]{Japanese\\restaurant} & \makecell[l]{Shabu-shabu\\restaurant} & -- \\
peking duck restaurant & 5 & Dining and Drinking & Restaurant & Asian restaurant & \makecell[l]{Chinese\\restaurant} & \makecell[l]{Peking duck\\restaurant} & -- \\
$\cdots$               & $\cdots$ & $\cdots$ & $\cdots$ & $\cdots$ & $\cdots$ & $\cdots$ & $\cdots$ \\
lake                   & 2 & Landmarks and Outdoors & Lake & -- & -- & -- & -- \\
barbershop             & 3 & \makecell[l]{Business and Professional\\Services} & \makecell[l]{Health and\\beauty service} & Barbershop & -- & -- & -- \\
hospice                & 2 & Health and Medicine & Hospice & -- & -- & -- & -- \\
\makecell[l]{Chinese aristocrat\\restaurant} & 5 & Dining and Drinking & Restaurant & Asian restaurant & \makecell[l]{Chinese\\restaurant} & \makecell[l]{Chinese aristocrat\\restaurant} & -- \\
\makecell[l]{Argentinian\\restaurant} & 5 & Dining and Drinking & Restaurant & \makecell[l]{Latin American\\restaurant} & \makecell[l]{South American\\restaurant} & \makecell[l]{Argentinian\\restaurant} & -- \\
\bottomrule
\end{tabular}
}
\caption{Hierarchical structure of the Foursquare taxonomy after cleaning and normalization, with categories organized across up to six depths.}
\label{fig:carte_fs}
\end{table*}

\subsubsection*{OpenStreetMap Dataset}
The OSM dataset was constructed from the official documentation using a custom scraping pipeline. Unlike FS, OSM does not provide a stable hierarchical taxonomy: POIs are instead defined by a flexible combination of keys and values that evolve continuously through community edits. The raw extraction therefore contained a heterogeneous set of elements, including attributes such as the number of building levels or construction year, which do not represent actual POIs and were consequently removed as deprecated or irrelevant tags (e.g., generic values such as \texttt{yes} or \texttt{user defined}) \cite{OSM_DeprecatedFeatures_2025}. Further cleaning focused on consolidating subcategories, since the OSM documentation distinguishes between headings outside and inside the value tables (\texttt{Subcategory\_before\_table} and \texttt{Subcategory\_in\_table}); these were merged into a single variable to avoid inconsistencies. Finally, we harmonized the taxonomy by removing underscores, standardizing depth levels, and restructuring the dataset into a format comparable to FS, with up to three levels of categorization (\texttt{Main category}, \texttt{Subcategory}, and \texttt{Value}).

After cleaning, the OSM taxonomy contains 1,205 valid tags distributed across 28 top-level categories. Unlike FS, these categories (e.g., \texttt{amenity}, \texttt{shop}, \texttt{leisure}) are not POI tags in themselves but rather entry points in the schema. Because our harmonization standardized depth levels such that each valid POI must be expressed at least as a key–value pair, no OSM tags terminate at depth~1. By contrast, FS includes actual categories already at this level.

Table~\ref{tab:osm_examples} shows the resulting structure, which remains shallower than FS (three levels versus six). Descriptive statistics comparing the two datasets are reported in Table~\ref{tab:osm-fs-summary}, highlighting both the number of top-level categories and the distribution of tags across depths.

\begin{table*}[htbp]
\centering
\footnotesize 
\setlength{\tabcolsep}{4pt} 
\renewcommand{\arraystretch}{0.9} 

\begin{tabularx}{\textwidth}{l c l l l X l}
\toprule
\textbf{Tag} & \textbf{Depth} & \textbf{Depth 1} & \textbf{Depth 2} & \textbf{Depth 3} & \textbf{Description} & \textbf{Element} \\
\midrule
cable car     & 2 & aerialway & cable car                   & --            & A cable car run; one or two large cars that shuttle along a line. & way \\
gondola       & 2 & aerialway & gondola                     & --            & An aerialway where enclosed cabins circulate continuously.        & way \\
mixed lift    & 2 & aerialway & mixed lift                  & --            & A mixed installation combining gondolas and chairlifts.           & way \\
chair lift    & 2 & aerialway & chair lift                  & --            & An open chairlift line with one or more seats per carrier.        & way \\
drag lift     & 2 & aerialway & drag lift                   & --            & An overhead tow-line intended for skiers and riders.              & way \\
$\cdots$      & $\cdots$ & $\cdots$ & $\cdots$ & $\cdots$ & $\cdots$ & $\cdots$ \\
lock gate     & 3 & waterway  & barriers on waterways       & lock gate     & A gate forming part of a navigation lock.                         & node / way \\
soakhole      & 3 & waterway  & other features on waterways & soakhole      & Point where a stream percolates into limestone or similar strata. & node \\
turning point & 3 & waterway  & other features on waterways & turning point & Location for vessels to reverse or change heading.                & node \\
water point   & 3 & waterway  & other features on waterways & water point   & Point to refill a boat’s fresh-water tanks.                       & node \\
fuel          & 3 & waterway  & other features on waterways & fuel          & Facility providing fuel for boats.                                & node / area \\
\bottomrule
\end{tabularx}
\caption{Examples from the OSM taxonomy (cleaned and harmonized): hierarchical context (Depth 1–3), textual description, and element type.}
\label{tab:osm_examples}
\end{table*}

\begin{table}[htbp]
  \centering
  \caption{Summary statistics for OSM and FS category datasets}
  \label{tab:osm-fs-summary}
  \begin{tabular}{l S[table-format=4.0] S[table-format=4.0]}
    \toprule
    & \multicolumn{1}{c}{\textbf{OSM}} & \multicolumn{1}{c}{\textbf{FS}} \\
    \midrule
    Total number of tags            & 1205 & 1244 \\
    Distinct top-level categories   & 28   & 11   \\
    \addlinespace[4pt]
    \multicolumn{3}{l}{\textit{Tag distribution by depth}} \\
    \midrule
    Depth 1 (top-level tags)        & \text{--}   & 11   \\
    Depth 2                         & 419  & 434  \\
    Depth 3                         & 786  & 464  \\
    Depth 4                         & \text{--}   & 239  \\
    Depth 5                         & \text{--}  & 82   \\
    Depth 6                         & \text{--}   & 14   \\
    \bottomrule
  \end{tabular}
\end{table}

\subsection*{Benchmark Construction}

\begin{figure}[h]
    \centering
    \includegraphics[width=.8\textwidth]{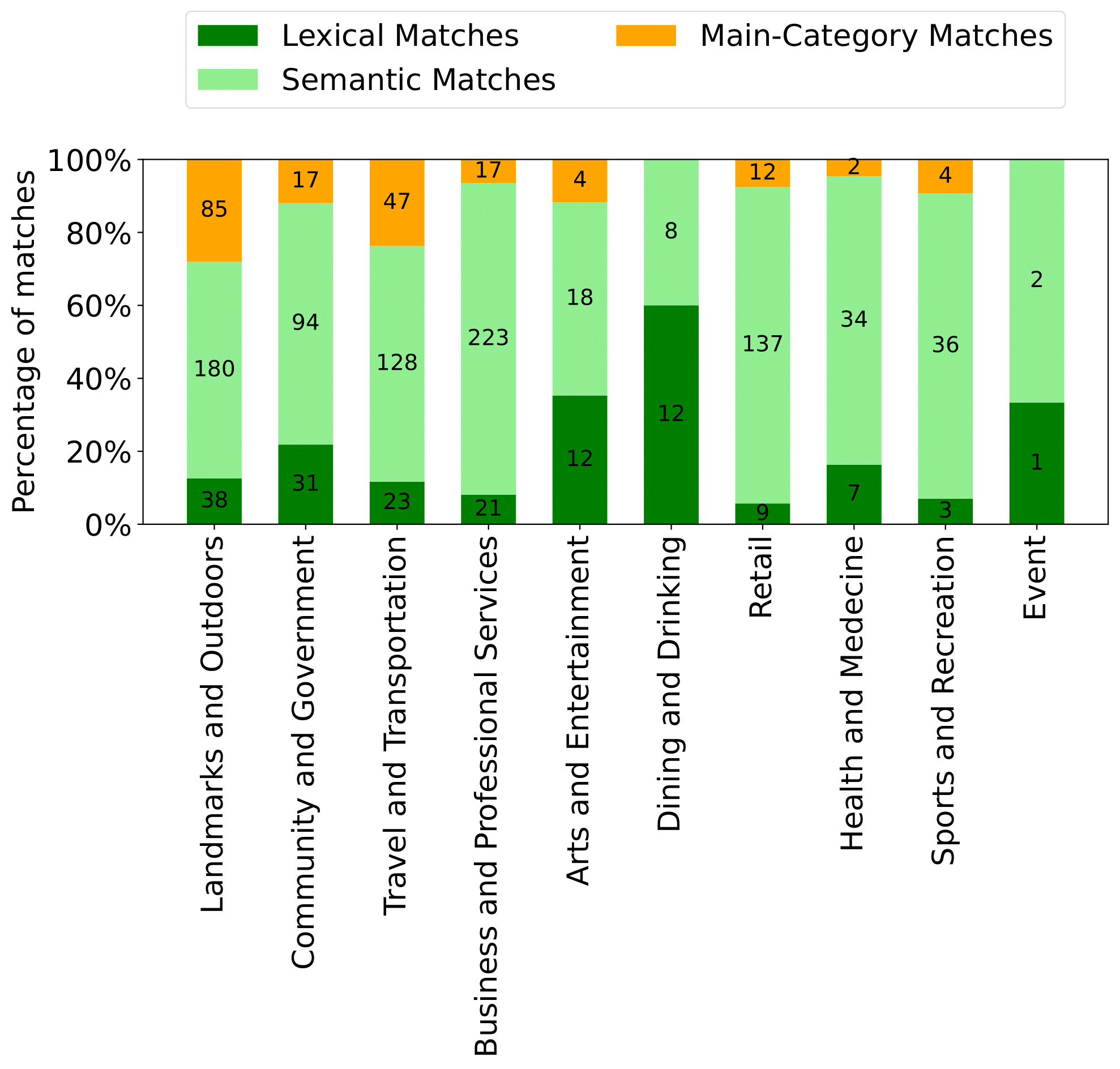}
    \caption{Absolute and relative frequencies of OSM–FS correspondences in the manual benchmark mapping, by FS main category and match type (lexical, semantic, main-category only).}
\label{fig:manual map}
\end{figure}

The construction of the benchmark starts from the two cleaned taxonomies introduced in the two previous subsections, which include 1,205 valid OSM tags and 1,244 FS categories. Establishing correspondences across such large and heterogeneous sets requires a structured process. 
A central component of our methodology is therefore the creation of a benchmark mapping in which OSM tags are aligned to FS categories, serving as the foundation for the quantitative evaluation of alignment methods. The benchmark combines three complementary types of correspondences.\\
The first type consists of exact lexical matches, where OSM tags and FS categories share the same label. These matches were nevertheless subject to manual verification, since identical labels may belong to different subcategories depending on their parent category, or may require disambiguation based on the detailed descriptions provided in OSM. After this validation, 157 OSM tags (approximately 13\% of the OSM set) were retained as reliable lexical matches. \\
The second type consists of semantically validated correspondences. Here, categories that do not share the same label but nevertheless denote equivalent concepts were aligned through manual inspection. OSM tags were compared to FS categories using their textual descriptions and hierarchical contexts. When a consistent semantic match could be established, including cases where FS aggregated multiple OSM variants under a broader label, the correspondence was included in the benchmark. This layer adds 860 OSM tags, corresponding to roughly 71\% of all matches, and extends coverage beyond trivial lexical overlap.\\
Finally, a third type includes OSM tags that cannot be mapped to any FS subcategory because no equivalent ones exist, but can still be associated with a broader FS main category. These cases capture structural mismatches between the two taxonomies and highlight systematic differences in granularity. Figure~\ref{fig:manual map} illustrates the distribution of these three types of correspondences across the FS main categories. \\
Analyzing the benchmark highlights systematic patterns. OSM tags with a perfect FS match are almost always located within subcategories, particularly in domains such as business services, retail, and leisure-related activities. These results confirm that FS is especially strong in covering urban environments and commercial services. Conversely, many OSM tags cannot be mapped at the same level of granularity and are absorbed into broad FS main categories. This is especially visible for domains such as geological features, natural elements, or barriers, which are richly represented in OSM but have no fine-grained counterpart in FS. In these cases, tags can only be placed into generic categories such as \texttt{Landmarks and Outdoors} or \texttt{Travel and Transportation}, underscoring structural mismatches between the two taxonomies. A complete list of the 188 OSM tags, grouped by OSM Depth~1 and their corresponding manual FS matches, is provided in Section~\ref{asec:tabmainmatch}(Tables~\ref{taba:13},\ref{taba:14},\ref{taba:15},\ref{taba:16},\ref{taba:17},\ref{taba:18},\ref{taba:19},\ref{taba:20},\ref{taba:21},\ref{taba:22},\ref{taba:23},\ref{taba:24},\ref{taba:25},\ref{taba:26},\ref{taba:27},\ref{taba:28},\ref{taba:29},\ref{taba:30},\ref{taba:31},\ref{taba:32},\ref{taba:33},\ref{taba:34}) of the Supplementary Information.
The benchmark, therefore, integrates lexical matches, semantically validated correspondences, and main-category assignments. It provides both a ground truth for the systematic evaluation of automatic alignment methods and an analytical lens on the semantic overlaps and gaps between OSM and FS. 
Openly sharing this curated and high-precision mapping between taxonomies aims at fostering urban science analyses at a more comprehensive level, improving transparency, and helping to reduce systematic and ad-hoc biases induced by partial and non-standardised POI mappings and category inclusion. 
The remaining part of this work focuses on making this dataset long-lived by providing code and analyses on the accuracy of future unsupervised integrations into the classification system \cite{osmfs_benchmark_2025}.

\subsection*{Sentence Embedding Models}
To perform the automatic alignment between OSM and FS categories, we experimented with eight pre-trained language models for text embeddings, all drawn from the Sentence-Transformers library \cite{reimers2019sentence}\footnote{\begin{minipage}[t]{0.95\linewidth}
The tested models are: 
\texttt{all-distilroberta-v1},
\texttt{all-MiniLM-L6-v2}, 
\texttt{all-MiniLM-L12-v2}, 
\texttt{all-mpnet-base-v2},  \texttt{paraphrase-distilroberta-base-v2},
\texttt{paraphrase-MiniLM-L6-v2}, 
\texttt{paraphrase-mpnet-base-v2},
and 
\texttt{paraphrase-tinybert-l6-v2.}
\end{minipage}}. 
The selection covers both lightweight and more resource-intensive architectures, as well as models optimized for either paraphrase detection or general-purpose similarity. Specifically, the lightweight group includes MiniLM variants \cite{wang2020minilm} and distilled RoBERTa models \cite{sanh2019distilbert,liu2019roberta}, and TinyBERT \cite{jiao2020tinybert}. The more resource-intensive group comprises MPNet variants \cite{song2020mpnet}. 
In more detail, MiniLM-based models \cite{wang2020minilm}, in their various configurations (\texttt{paraphrase-MiniLM-L6-v2}, \texttt{all-MiniLM-L6-v2}, and \texttt{all-MiniLM-L12-v2}), are designed to be compact and efficient, making them well suited for large-scale applications. MPNet models \cite{song2020mpnet}, represented here by \texttt{all-mpnet-base-v2} and \texttt{paraphrase-mpnet-base-v2}, consistently achieve state-of-the-art performance in semantic similarity tasks; the paraphrase-oriented variant is particularly effective for fine-grained category alignment. DistilRoBERTa \cite{sanh2019distilbert}, tested in both its all-purpose (\texttt{all-distilroberta-v1}) and paraphrase-specialized (\texttt{paraphrase-distilroberta-base-v2}) forms, offers a compromise between accuracy and computational efficiency, retaining much of the expressiveness of RoBERTa while reducing its size. Finally, \texttt{paraphrase-tinybert-l6-v2} \cite{jiao2020tinybert} provides an extremely compact alternative that enables experimentation under constrained computational resources, albeit at the cost of some loss in accuracy. In all cases, the \texttt{paraphrase-*} series is trained for sentence-level entailment and paraphrase identification, whereas the \texttt{all-*} series is designed for general-purpose semantic similarity \cite{reimers2019sentence}.
This diversity enables us to investigate how different embedding families capture semantic relations between heterogeneous taxonomies and to assess their robustness with respect to the evolving nature of OSM tags.\\
All models were tested using the textual descriptions associated with OSM tags provided in the dataset. 
For FS categories, we adopted two alternative representations: one relying only on the official category labels, and another enriched with concise textual descriptions automatically generated with ChatGPT-5 (OpenAI \cite{openai_chatgpt5}).
The generation was performed through batch prompting of the web interface, using a controlled instruction of the form: 
“Here is a list of Foursquare points of interest. For each one, give me a simple and precise description (about one sentence) of what the tag represents.”. 
This procedure produced short, consistent definitions that were added to the \texttt{def\_tags\_fs.csv} file available in the public repository \cite{osmfs_benchmark_2025}. 
This design allows us to assess whether additional descriptive context improves the ability of embedding models to capture cross-taxonomy semantic relations.\\
To evaluate alignments before the refinement stage, we compute cosine similarity scores between the embeddings of each OSM tag and all FS categories \cite{mardia1979multivariate}. Since similarity ranges differ substantially across tags, we do not use a global threshold. Instead, we focus on the top-scoring FS category (top-$1$) for each OSM tag. A match is considered reached if the retrieved FS category path attains at least the depth level established in the manual benchmark. This criterion accounts for cases where multiple FS subcategories provide equally plausible synonyms of an OSM tag, while the benchmark conservatively stopped at a broader category. For example, the OSM tag \texttt{valley} could correspond to multiple FS subcategories, such as \texttt{Landmarks and Outdoors > Hill}, \texttt{Landmarks and Outdoors > Mountain}, or \texttt{Landmarks and Outdoors > Other Great Outdoors}. In such cases, the match is validated as long as the model retrieves at least the broader parent category \texttt{Landmarks and Outdoors}.\\
From these per-tag outcomes, we derive two complementary metrics. 
Alignment accuracy quantifies the proportion of OSM tags whose top-ranked FS category matches the benchmark mapping, corresponding to top-1 accuracy \cite{sokolova2006systematic}. 
We further report top-k recall (for $k \in \{5, 10, 20\}$), defined as the proportion of benchmark matches retrieved within the top-k most similar candidates. 
For simplicity, we refer to both measures as top-k metrics, with top-1 representing accuracy and higher k values representing recall. In addition, ROC-AUC \cite{fawcett2006roc} (Receiver Operating Characteristic - Area Under the Curve), a standard threshold-independent measure of discriminative ability, evaluates how well similarity scores separate correct from incorrect candidates across all thresholds, offering a global assessment of ranking quality. Considering these metrics together allows us to capture both direct retrieval accuracy and the robustness of candidate rankings that underpin the refinement process.


\subsection*{LLM-based Refinement}\label{subsec:meth_ref}

The final step of our methodology builds on the candidate alignments produced by the embedding models. For each OSM tag, we consider the top-$k$ FS categories with the highest similarity scores ($k \in \{5, 10, 20, 30, 40, 50\}$). These candidate categories are then provided as input to \texttt{ChatGPT-4o-mini} via the OpenAI API \cite{openai_gpt4omini}, guided by prompts detailed in Supplementary Information \ref{asec:prompt}) that ask the model to identify the most semantically appropriate mapping. We also tested performances with the \texttt{ChatGPT-5-mini} \cite{openai_gpt5mini} model, finding no substantial improvement and, on the contrary, burdening the pipeline with considerably more computational effort, thus reducing efficiency. \\
The introduction of this refinement layer is motivated by the observation that embedding-based retrieval, while capturing broad semantic similarity, often fails to discriminate between subtle but meaningful distinctions. By combining embeddings with LLM-based re-ranking, we aim to enhance alignment accuracy and ensure robustness against the continuous evolution of OSM tags, where new or ambiguous entries can emerge. In this way, the framework integrates the scalability of dense similarity search with the semantic depth of generative models, providing a more resilient solution for POI taxonomy alignment. \\
To assess the impact of prompt design, we tested four prompting strategies varying along two dimensions: the use of fallback to broad FS categories and the inclusion of a worked example. The full prompt formulations are provided in Supplementary Information \ref{asec:prompt}.\\
Unlike the embedding-based retrieval stage, the refinement process outputs only a single selected candidate without an associated similarity score. To ensure comparability with the previous evaluation framework, we assign the chosen candidate a surrogate score equal to the maximum similarity among the retrieved top-$k$ candidates, incremented by a small $\epsilon$ for tie-breaking. This procedure, which we refer to as \emph{max-score imputation with $\epsilon$ adjustment}, preserves consistency across evaluation stages while reflecting the refinement’s decisive choice.

\section*{Data Records}

All processed datasets, benchmark mappings, and cleaned category tables generated in this study are publicly released in the GitHub repository Mapping-of-OSM-and-FS-categories \cite{osmfs_benchmark_2025}. The repository contains the following data records and artifacts:\\
{\bf Cleaned category datasets:} \\ 
Two CSV files, `categories\_OSM\_clean.csv' and `categories\_FS\_clean.csv', containing the cleaned OSM and Foursquare taxonomy categories, respectively, after normalization and filtering.  \\
{\bf Enriched FS dataset with descriptions:}  \\
A CSV file `categories\_FS\_clean\_description.csv', which augments the FS categories with descriptive labels integrated via ChatGPT prompts.\\
{\bf Benchmark (oracle) mapping:}  \\
  A CSV file `df\_oracle.csv' linking each OSM tag in the benchmark to its manually curated FS counterpart, with metadata on match type (lexical, semantic, structural).\\
{\bf Embedding \& alignment results:}  \\
  Output files (e.g. CSVs) from the embedding and taxonomy alignment steps, stored under directories such as `FID gpt models' and `Embedding' within the repository, corresponding to model predictions and performance metrics across top-$k$ thresholds.\\
{\bf Jupyter notebooks:}  \\
  Seven notebooks (e.g. `01-download\_OSM\_map\_features\_tables.ipynb',\\
  `02-categories\_OSM\_FS\_datasets\_cleaning.ipynb', up to `07-final\_comparaison.ipynb') that reproduce the full data pipeline: data acquisition, cleaning, embedding, benchmark generation, model prediction, and the generation of tables and figures.  \\
All these data and code artifacts are versioned in the repository and are retrievable via Git. 
The repository is the primary access point for the data outputs described in this manuscript, and users may download individual files or clone the full repository for reproducibility.  \\

\section*{Technical Validation}
This section reports the results obtained in evaluating the proposed alignment framework. The evaluation relies on the benchmark mapping, using best-match accuracy and ROC-AUC as metrics. We first analyze the coverage of FS categories in the benchmark. We then compare embeddings with and without FS category descriptions, and finally present the results of the embedding-based retrieval and the LLM-based refinement.

\subsection*{Coverage of FS Categories in the Benchmark}

Before assessing automatic methods, we first analyze the coverage of FS categories within the manually constructed benchmark. Table~\ref{tab:diff_FS} reports the number of distinct FS tags that were actually used to classify OSM tags, compared to the total number of categories available in FS. Only 463 distinct FS category paths out of 1244 are represented in the benchmark, indicating that the overlap between the two taxonomies is concentrated in a limited subset of FS categories.\\
The distribution is highly uneven across main categories. For instance, 82\% of the categories in \texttt{Landmarks and Outdoors}, nearly 70\% of those in \texttt{Travel and Transportation}, and about two-thirds of those in \texttt{Retail} are represented.
However, many of these correspondences involve OSM tags without a perfect match that are aggregated under broad FS labels, consistent with the structural mismatches between the two taxonomies. Conversely, categories such as \texttt{Dining and Drinking} or \texttt{Nightlife Spot} are scarcely used, reflecting the relative scarcity of such tags in OSM compared to FS. These results provide a first quantitative characterization of the benchmark and establish a reference point for evaluating the performance of automatic alignment methods.
\begin{table}[htbp]
  \centering
  \caption{Coverage of FS main categories in the manual benchmark.}
  \label{tab:diff_FS}
  \begin{tabular}{l 
                  S[table-format=3.0] 
                  S[table-format=3.0] 
                  S[table-format=2.2]}
    \toprule
    \textbf{Main FS category} 
      & {\begin{tabular}[c]{@{}c@{}}Nr. of FS\\ categories\\ used\end{tabular}} 
      & {\begin{tabular}[c]{@{}c@{}}Nr. of FS\\ categories\\ available\end{tabular}} 
      & {\begin{tabular}[c]{@{}c@{}}Coverage\\ (\%)\end{tabular}} \\
    \midrule
    \makecell{\texttt{Landmarks}\\ \texttt{and Outdoors}}            & 58  & 71  & 81.69 \\
    \addlinespace[4pt]
    \makecell{\texttt{Travel and}\\ \texttt{Transportation}}         & 49  & 72  & 68.06 \\
    \addlinespace[4pt]
    \makecell{\texttt{Retail}}                                       & 101 & 150 & 67.33 \\
    \addlinespace[4pt]
    \makecell{\texttt{Business and}\\ \texttt{Professional}\\ \texttt{Services}} & 112 & 195 & 57.44 \\
    \addlinespace[4pt]
    \makecell{\texttt{Community and}\\ \texttt{Government}}          & 53  & 127 & 41.73 \\
    \addlinespace[4pt]
    \makecell{\texttt{Health and}\\ \texttt{Medicine}}               & 22  & 59  & 37.29 \\
    \addlinespace[4pt]
    \makecell{\texttt{Arts and}\\ \texttt{Entertainment}}            & 26  & 72  & 36.11 \\
    \addlinespace[4pt]
    \makecell{\texttt{Event}}                                        & 3   & 17  & 17.65 \\
    \addlinespace[4pt]
    \makecell{\texttt{Sports and}\\ \texttt{Recreation}}             & 23  & 87  & 26.44 \\
    \addlinespace[4pt]
    \makecell{\texttt{Dining and}\\ \texttt{Drinking}}               & 17  & 392 & 4.33  \\
    \addlinespace[4pt]
    \makecell{\texttt{Nightlife Spot}}                               & 0   & 2   & 0.00  \\
    \bottomrule
  \end{tabular}
\end{table}

\subsection*{Effect of FS Descriptions on Embedding Models}
\begin{figure}[htbp]
  \centering
  \begin{subfigure}{0.48\textwidth}
    \centering
    \includegraphics[width=\linewidth]{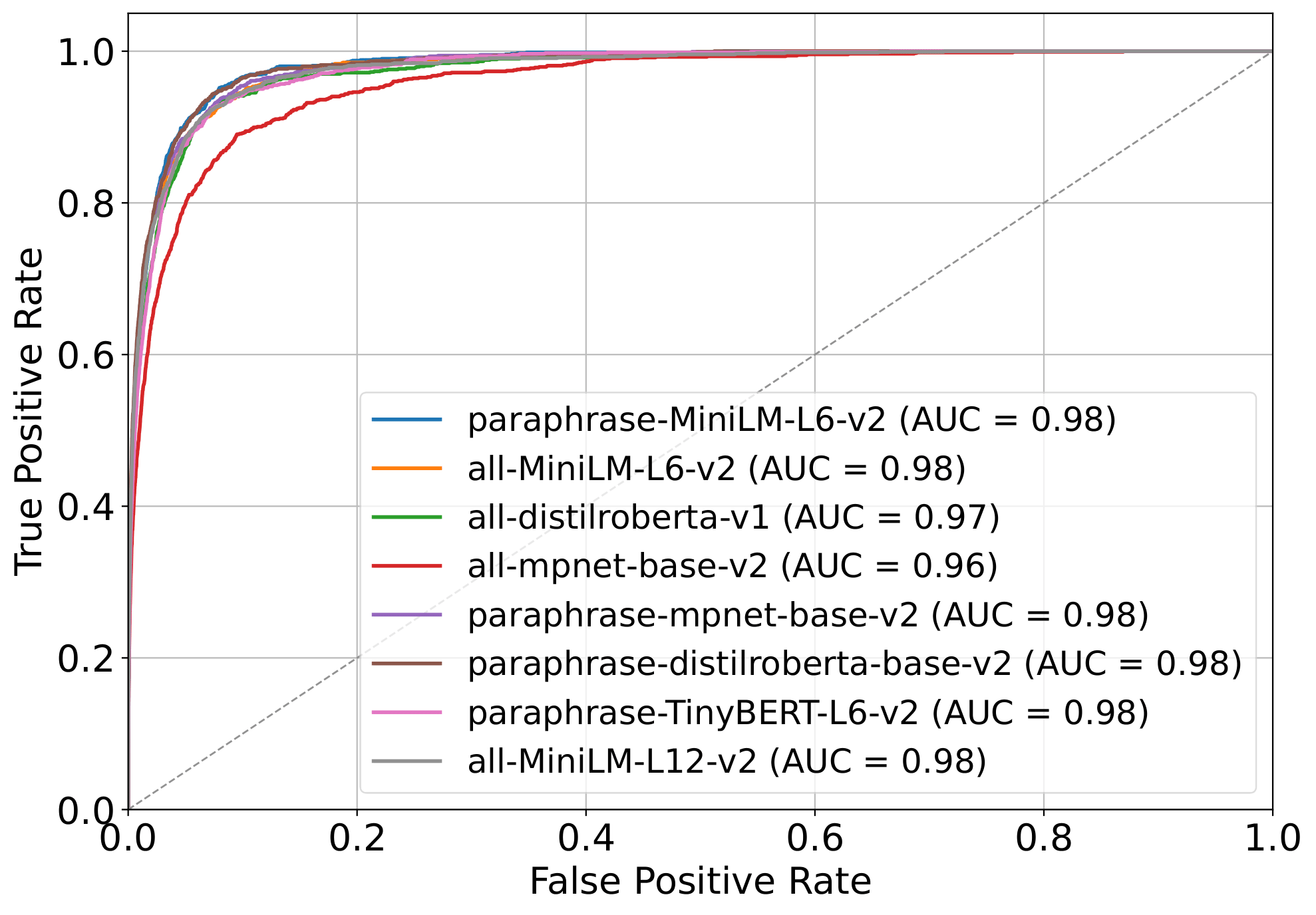}
    \caption{Without FS descriptions (FI)}
    \label{fig:roc_fi}
  \end{subfigure}
  \hfill
  \begin{subfigure}{0.48\textwidth}
    \centering
    \includegraphics[width=\linewidth]{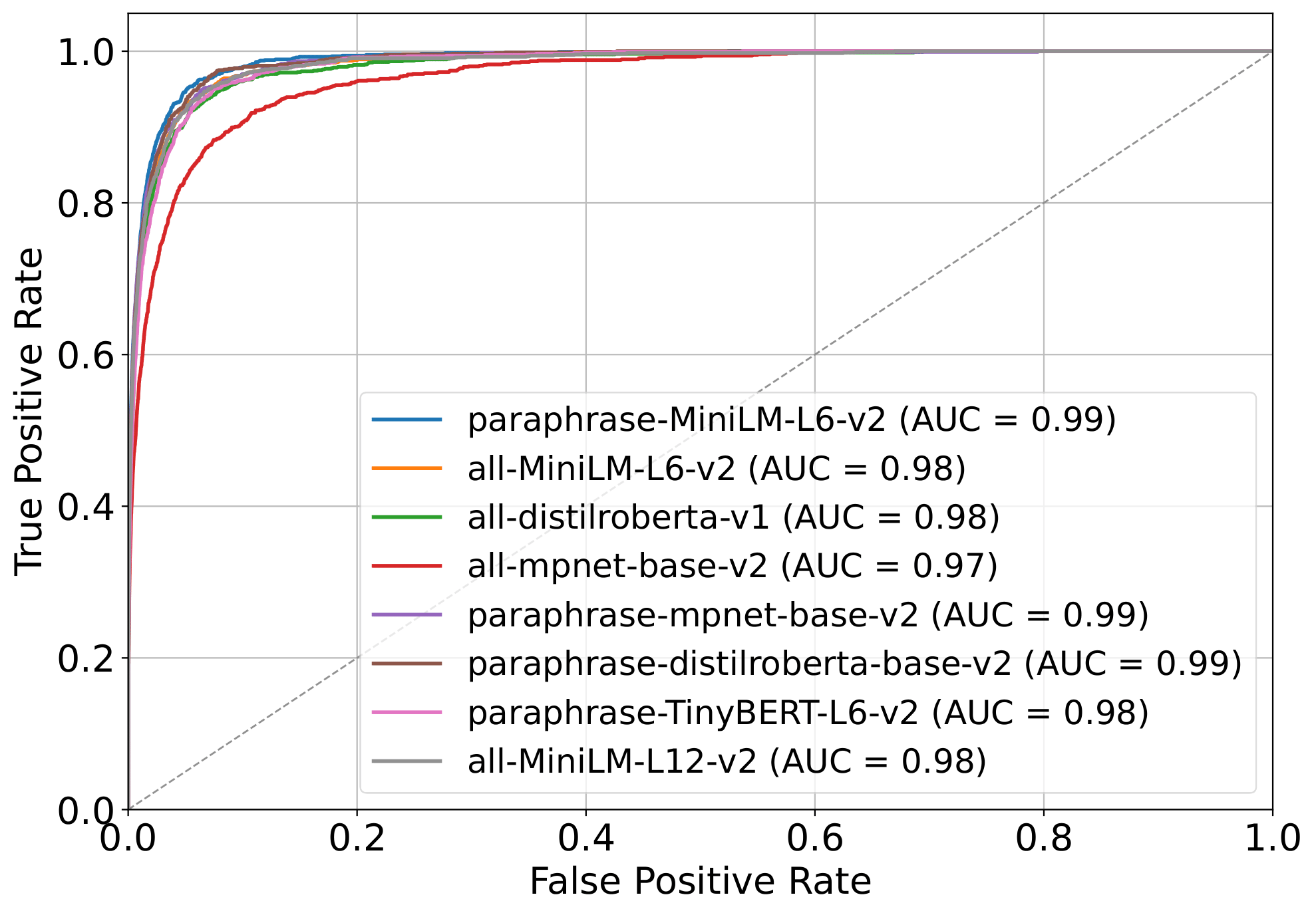}
    \caption{With FS descriptions (FID)}
    \label{fig:roc_fid}
  \end{subfigure}
  \caption{ROC curves for embedding-based retrieval with and without FS category descriptions. 
  Adding descriptions improves the discriminative ability of similarity scores, even when top-1 accuracy slightly decreases.}
  \label{fig:roc_fi_fid}
\end{figure}

Table~\ref{tab:fs_description_comparison} and Figure~\ref{fig:roc_fi_fid} jointly summarize the effect of enriching FS categories with textual descriptions. The ROC curves show very high discriminative performance, with AUCs above 0.97 without descriptions and above 0.98 with descriptions, except for \texttt{all-mpnet-base-v2}, which remains slightly lower (0.96 and 0.97, respectively). The inclusion of textual descriptions increases ROC--AUC for nearly all models by about 0.6--0.7\%, confirming that enrichment systematically improves ranking quality. By contrast, top-1 accuracy exhibits mixed behavior, improving for some models while decreasing for others, indicating that robustness in ranking does not always coincide with immediate retrieval.  \\
Additionally, we evaluated the top-1 accuracy across different depths of the hierarchy using models with enriched FS category descriptions. Notably, while all models achieve over 78\% accuracy at Depth 1 (with the exception of \texttt{all-mpnet-base-v2}, which still reaches a respectable 76\%), the performance across the remaining depths varies, showing the models' ability to adapt to more granular categorization. Among the models, the \texttt{all-MiniLM} and \texttt{all-distilroberta} models perform among the top choices, maintaining the best accuracy values across the different depths. These results are shown in Table~\ref{tab:model_performance_comparison}.

\begin{table*}[htbp]
  \centering
  \caption{Comparison of model performance with and without FS category descriptions. 
  FI = models evaluated using only FS labels; 
  FID = models evaluated using FS labels enriched with descriptions. 
  Metrics refer to the evaluation setting.}
  \label{tab:fs_description_comparison}
  \begin{tabular}{l 
                  S[table-format=2.2] S[table-format=2.2] S[table-format=+1.2] 
                  S[table-format=1.3] S[table-format=1.3] S[table-format=+1.3]}
    \toprule
    & \multicolumn{3}{c}{\textbf{Top-1 acc. (\%)}} 
    & \multicolumn{3}{c}{\textbf{ROC-AUC}} \\
    \cmidrule(lr){2-4} \cmidrule(lr){5-7}
    \textbf{Model} 
      & {FI} & {FID} & {$\Delta$} 
      & {FI} & {FID} & {$\Delta$} \\
    \midrule
    all-distilroberta-v1             & 56.93 & 56.93 &  0.00 & 0.974 & 0.981 & +0.007 \\
    all-MiniLM-L6-v2                 & 58.84 & 57.18 & -1.66 & 0.978 & 0.984 & +0.006 \\
    all-MiniLM-L12-v2                & 56.18 & 57.01 & +0.83 & 0.976 & 0.983 & +0.007 \\
    all-mpnet-base-v2                & 51.70 & 52.86 & +1.16 & 0.959 & 0.966 & +0.007 \\
    paraphrase-distilroberta-base-v2 & 56.18 & 55.35 & -0.83 & 0.981 & 0.987 & +0.006 \\
    paraphrase-MiniLM-L6-v2          & 53.69 & 55.68 & +1.99 & 0.981 & 0.988 & +0.007 \\
    paraphrase-mpnet-base-v2         & 57.84 & 54.19 & -3.65 & 0.979 & 0.985 & +0.006 \\
    paraphrase-TinyBERT-L6-v2        & 53.28 & 54.11 & +0.83 & 0.976 & 0.982 & +0.006 \\
    \bottomrule
  \end{tabular}
\end{table*}

\begin{table*}[htbp]
  \centering
  
  \setlength{\tabcolsep}{4pt}
  \renewcommand{\arraystretch}{0.9}
  \caption{Model performance comparison across different depths with enriched FS category descriptions.}
  \label{tab:model_performance_comparison}
  \begin{tabularx}{\textwidth}{l *{6}{S[table-format=2.2]}}
    \toprule
    \textbf{Model} & \multicolumn{6}{c}{\textbf{Top-1 acc. (\%)}} \\
    \cmidrule(lr){2-7}
    & \textbf{Depth 1} & \textbf{Depth 2} & \textbf{Depth 3} & \textbf{Depth 4} & \textbf{Depth 5} & \textbf{Depth 6} \\
    \midrule
    all-distilroberta-v1                 & 78.67 & 62.24 & 57.34 & 56.93 & 56.93 & 56.93 \\
    all-MiniLM-L6-v2                     & 78.59 & 61.99 & 57.76 & 57.18 & 57.18 & 57.18 \\
    all-MiniLM-L12-v2                    & 78.67 & 61.74 & 57.01 & 57.01 & 57.01 & 57.01 \\
    all-mpnet-base-v2                    & 76.27 & 57.10 & 53.03 & 52.86 & 52.86 & 52.86 \\
    paraphrase-distilroberta-base-v2     & 78.34 & 60.58 & 55.77 & 55.35 & 55.35 & 55.35 \\
    paraphrase-MiniLM-L6-v2              & 78.09 & 61.00 & 56.02 & 55.68 & 55.68 & 55.68 \\
    paraphrase-mpnet-base-v2             & 78.26 & 59.75 & 54.77 & 54.19 & 54.19 & 54.19 \\
    paraphrase-TinyBERT-L6-v2            & 78.67 & 58.76 & 54.36 & 54.11 & 54.11 & 54.11 \\
    \bottomrule
  \end{tabularx}
\end{table*}

Table~\ref{tab:fs_topk_comparison} extends the analysis to Top-5, Top-10, and Top-20 recall, while Table~\ref{tab:fs_topk_comparison_large} reports results up to Top-50. Across all thresholds, the inclusion of textual descriptions leads to systematic improvements in top-$k$ recall. The best percentages, for every $k$, are typically achieved by the \texttt{all-MiniLM} family, followed by \texttt{paraphrase-distilroberta}, \texttt{MiniLM}, and \texttt{mpnet}. The advantage of \texttt{all-MiniLM} is most pronounced at lower cutoffs (up to $k=20$), while for larger shortlists ($k\geq30$) performance tends to converge across models.\\  
In selecting a model for downstream refinement, we prioritize ROC–AUC as the most reliable indicator of ranking quality, and we complement it with top-$k$ recall to assess the quality of the candidate sets that will be passed to the generative model. Overall, the \texttt{all-MiniLM} family remains the most competitive, combining robust AUC values with consistently high top-$k$ scores. Importantly, the choice to rely on \texttt{all-MiniLM} is also motivated by its higher efficiency compared to the other candidates, making it the most suitable option for subsequent experiments \cite{sbert_pretrained_models}. It is important to note, however, that higher top-$k$ values do not necessarily translate into better recall of the final answer: enlarging the shortlist increases the number of proposals presented to ChatGPT, which may reduce its ability to select the correct one. Therefore, careful consideration is needed to identify the best trade-off between ranking robustness and shortlist size.\\
For the subsequent refinement stage, we therefore adopt the \texttt{all-MiniLM-L6-v2} model with enriched FS category descriptions, as it combines the best trade-off between accuracy, robustness, and efficiency.

\begin{table*}[htbp]
  \centering
  \setlength{\tabcolsep}{4pt} 
  \renewcommand{\arraystretch}{0.9} 
  \caption{Comparison of model performance with and without FS category descriptions across Top-$k$ recall. 
  FI = models evaluated using only FS labels; 
  FID = models evaluated using FS labels enriched with descriptions.}
  \label{tab:fs_topk_comparison}
  \begin{tabularx}{\textwidth}{
    l
                  S[table-format=2.2] S[table-format=2.2] S[table-format=+1.2]
                  S[table-format=2.2] S[table-format=2.2] S[table-format=+1.2]
                  S[table-format=2.2] S[table-format=2.2] S[table-format=+1.2]}
    \toprule
    & \multicolumn{3}{c}{\textbf{Top-5 recall (\%)}} 
    & \multicolumn{3}{c}{\textbf{Top-10 recall (\%)}} 
    & \multicolumn{3}{c}{\textbf{Top-20 recall (\%)}} \\
    \cmidrule(lr){2-4}\cmidrule(lr){5-7}\cmidrule(lr){8-10}
    \textbf{Model} 
      & {FI} & {FID} & {$\Delta$}
      & {FI} & {FID} & {$\Delta$}
      & {FI} & {FID} & {$\Delta$} \\
    \midrule
    all-distilroberta-v1             & 81.66 & 84.15 & +2.49 & 88.46 & 90.62 & +2.16 & 93.94 & 94.44 & +0.50 \\
    all-MiniLM-L6-v2                 & 85.06 & 86.39 & +1.33 & 90.79 & 92.37 & +1.58 & 94.11 & 95.44 & +1.33 \\
    all-MiniLM-L12-v2                & 84.98 & 85.64 & +0.66 & 90.79 & 91.78 & +0.99 & 94.44 & 95.60 & +1.16 \\
    all-mpnet-base-v2                & 80.08 & 81.66 & +1.58 & 86.47 & 89.05 & +2.57 & 92.03 & 93.69 & +1.66 \\
    paraphrase-distilroberta-base-v2 & 81.16 & 83.40 & +2.24 & 87.81 & 89.96 & +2.15 & 93.03 & 95.19 & +2.16 \\
    paraphrase-MiniLM-L6-v2          & 81.24 & 83.07 & +1.83 & 88.05 & 90.12 & +2.07 & 91.87 & 94.85 & +2.99 \\
    paraphrase-mpnet-base-v2         & 83.07 & 84.98 & +1.91 & 88.46 & 90.79 & +2.32 & 93.20 & 94.11 & +0.91 \\
    paraphrase-TinyBERT-L6-v2        & 79.25 & 81.83 & +2.57 & 86.72 & 89.38 & +2.66 & 91.70 & 93.94 & +2.24 \\
    \bottomrule
  \end{tabularx}
\end{table*}

\begin{table*}[htbp]
  \centering
  \setlength{\tabcolsep}{4pt} 
  \renewcommand{\arraystretch}{0.9} 
  \caption{Comparison of model performance with and without FS category descriptions across Top-$k$ recall for larger $k$ values.}
  \label{tab:fs_topk_comparison_large}
  \begin{tabular}{l
                  S[table-format=2.2] S[table-format=2.2] S[table-format=+1.2]
                  S[table-format=2.2] S[table-format=2.2] S[table-format=+1.2]
                  S[table-format=2.2] S[table-format=2.2] S[table-format=+1.2]}
    \toprule
    & \multicolumn{3}{c}{\textbf{Top-30 recall (\%)}} 
    & \multicolumn{3}{c}{\textbf{Top-40 recall (\%)}} 
    & \multicolumn{3}{c}{\textbf{Top-50 recall (\%)}} \\
    \cmidrule(lr){2-4}\cmidrule(lr){5-7}\cmidrule(lr){8-10}
    \textbf{Model} 
      & {FI} & {FID} & {$\Delta$}
      & {FI} & {FID} & {$\Delta$}
      & {FI} & {FID} & {$\Delta$} \\
    \midrule
    all-distilroberta-v1             & 95.35 & 96.02 & +0.66 & 95.93 & 96.76 & +0.83 & 96.43 & 97.26 & +0.83 \\
    all-MiniLM-L6-v2                 & 96.10 & 96.43 & +0.33 & 96.60 & 97.10 & +0.50 & 97.01 & 97.26 & +0.25 \\
    all-MiniLM-L12-v2                & 95.77 & 96.76 & +1.00 & 96.51 & 97.34 & +0.83 & 97.01 & 97.84 & +0.83 \\
    all-mpnet-base-v2                & 94.19 & 95.44 & +1.25 & 95.10 & 96.18 & +1.08 & 96.10 & 96.60 & +0.50 \\
    paraphrase-distilroberta-base-v2 & 95.35 & 96.08 & +0.73 & 96.51 & 97.43 & +0.91 & 96.85 & 98.17 & +1.33 \\
    paraphrase-MiniLM-L6-v2          & 94.36 & 97.01 & +2.66 & 96.01 & 97.59 & +1.58 & 96.76 & 97.84 & +1.08 \\
    paraphrase-mpnet-base-v2         & 95.52 & 95.85 & +0.33 & 96.60 & 96.93 & +0.33 & 97.26 & 97.59 & +0.33 \\
    paraphrase-TinyBERT-L6-v2        & 94.27 & 95.35 & +1.08 & 95.44 & 96.18 & +0.75 & 96.76 & 97.10 & +0.33 \\
    \bottomrule
  \end{tabular}
\end{table*}

\subsection*{Evaluation of Refinement Results}
In this subsection, we assess the impact of the LLM-based refinement procedure. Starting from the candidate alignments retrieved by the \texttt{all-MiniLM-L6-v2} model, enriched with the textual descriptions of FS categories, we tested four prompting strategies across six values of $k$. These strategies combine two factors: (i) whether the model could fall back to a broad first-level FS category when none of the top-$k$ candidates provided an adequate match, and (ii) whether a worked example was included in the instruction. Table~\ref{tab:prompt_summary} provides a summary of these configurations, while the complete prompt formulations are reported in Supplementary Information ~\ref{asec:prompt}.
As observed in Table~\ref{tab:fs_topk_comparison} and Table~\ref{tab:fs_topk_comparison_large}, increasing $k$ systematically enlarges the fraction of OSM tags whose manual match is included among the retrieved candidates. However, this does not necessarily imply a higher probability of selecting the correct match. Indeed, across all prompting strategies, recall tends to decrease when $k$ grows beyond 20. For smaller values of $k$ (i.e., 5, 10, and 20), the results show more variability across prompts and hierarchical depths. For example, at the second and third depth levels, prompt~\texttt{No fallback - No example} and prompt~\texttt{Fallback - No example} benefit from larger candidate sets, leading to slight improvements as $k$ increases. Table~\ref{tab:model_performance_comparison_k} reports the results of all prompts across values of $k$ and depth levels.\\
\begin{table}[ht]
\centering
\caption{Summary of the four prompting strategies used for OSM–FS category mapping.}
\label{tab:prompt_summary}
\begin{tabular}{ll}
\toprule
\textbf{Prompt}  & \textbf{Description} \\
\midrule
No fallback – No example & Direct matching based on rules; no fallback categories or example. \\
Fallback – No example &  Same as above, but allows fallback to 10 broad FS categories. \\
No fallback – With example &  Includes a worked example; no fallback categories. \\
Fallback – With example & Combines both fallback mechanism and example-based instruction. \\
\bottomrule
\end{tabular}
\end{table}

\begin{figure}[htbp]
  \centering
      \begin{subfigure}[]{0.4\textwidth}
      \vspace{0pt}
    \centering
    \includegraphics[width=\linewidth]{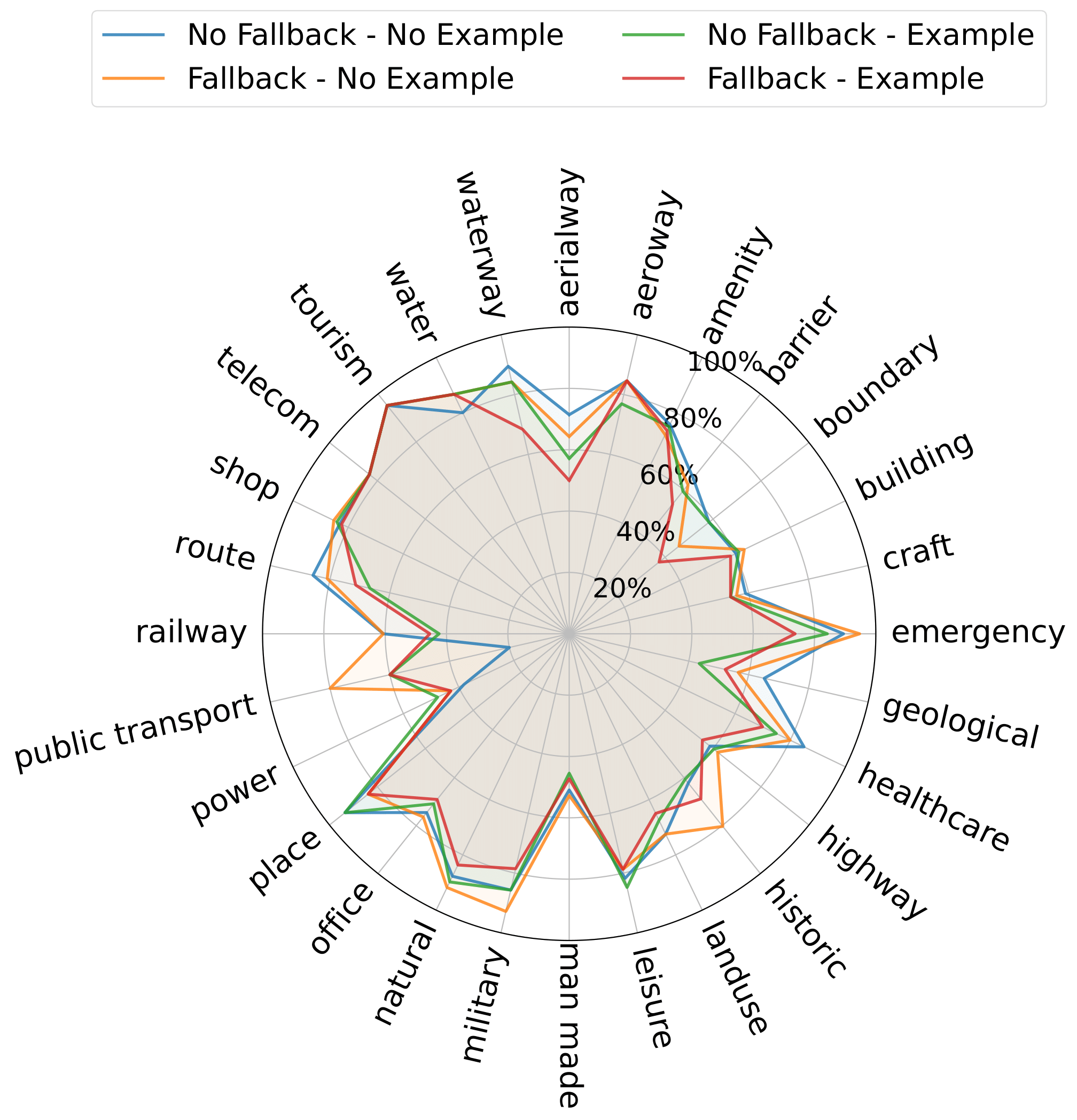}
    \caption{}
    \label{fig:k20_osm}
  \end{subfigure}
    \begin{subfigure}[]{0.5\textwidth}
    \vspace{0pt}
    \centering
    \includegraphics[width=\linewidth]{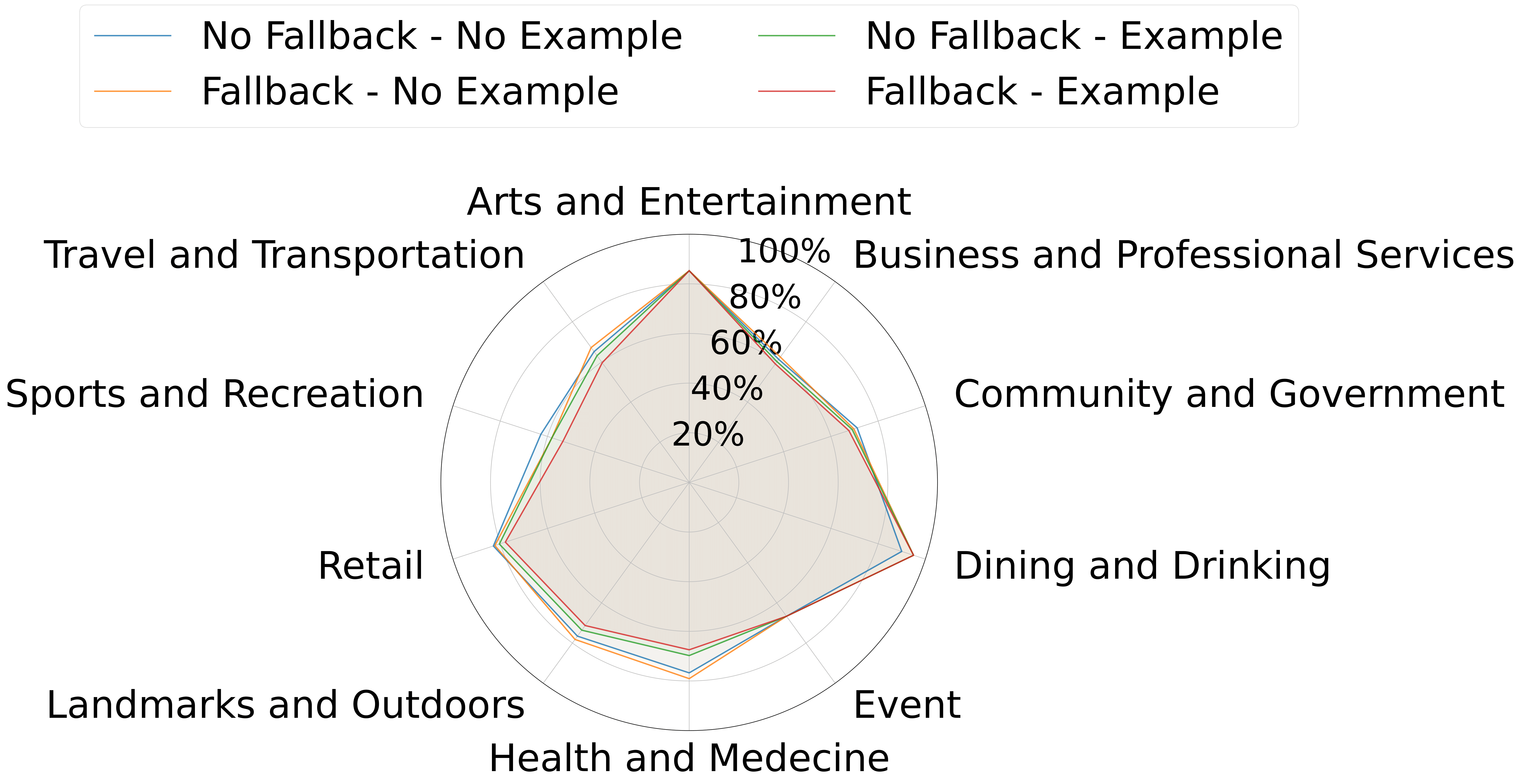}
    \caption{}
    \label{fig:k20_fs}
  \end{subfigure}
          \caption{Comparison of match percentages for OSM and FS categories at $k=20$ across the four prompting strategies. On the left, the radar chart shows the match percentages across OSM tags, while on the right, the radar chart displays the corresponding match percentages for FS categories.}
  
  \label{fig:k_osm_fs}
\end{figure}

\begin{figure}[htbp]
  \centering
   \includegraphics[width=0.8\linewidth]{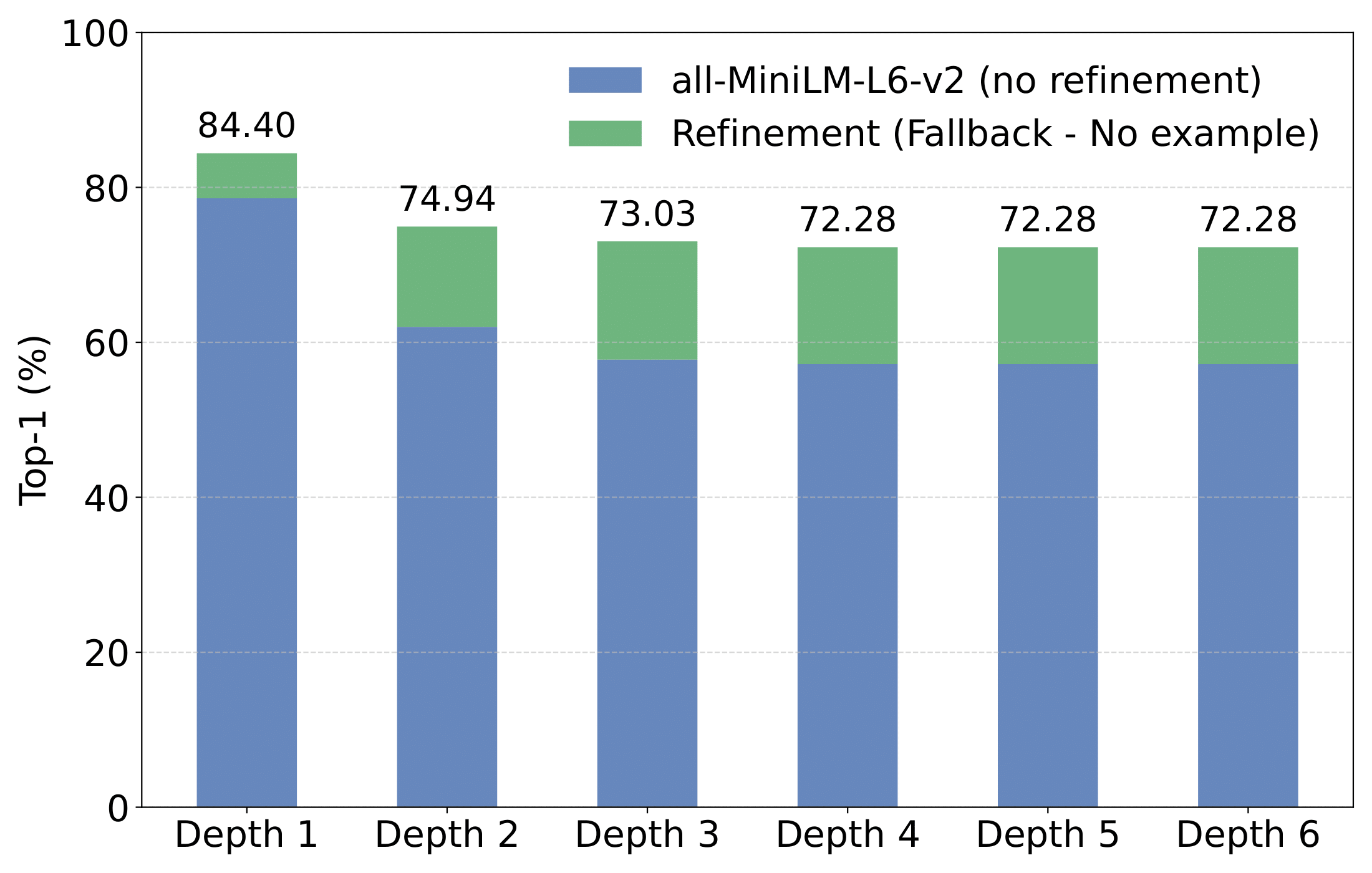}
    \caption{Top-1 accuracy (\%) by hierarchy depth for the \texttt{all-MiniLM-L6-v2} model. Each bar shows the baseline performance without refinement (blue) and the additional improvement achieved with the \texttt{Fallback – No example} prompt at $k=20$ (green). Values are annotated at the top of each bar. }
  \label{fig:mdel_ref}
\end{figure}

\begin{figure}[htbp]
  \centering
    \begin{subfigure}[]{0.4\textwidth}
    \vspace{0pt}
    \centering
    \includegraphics[width=\linewidth]{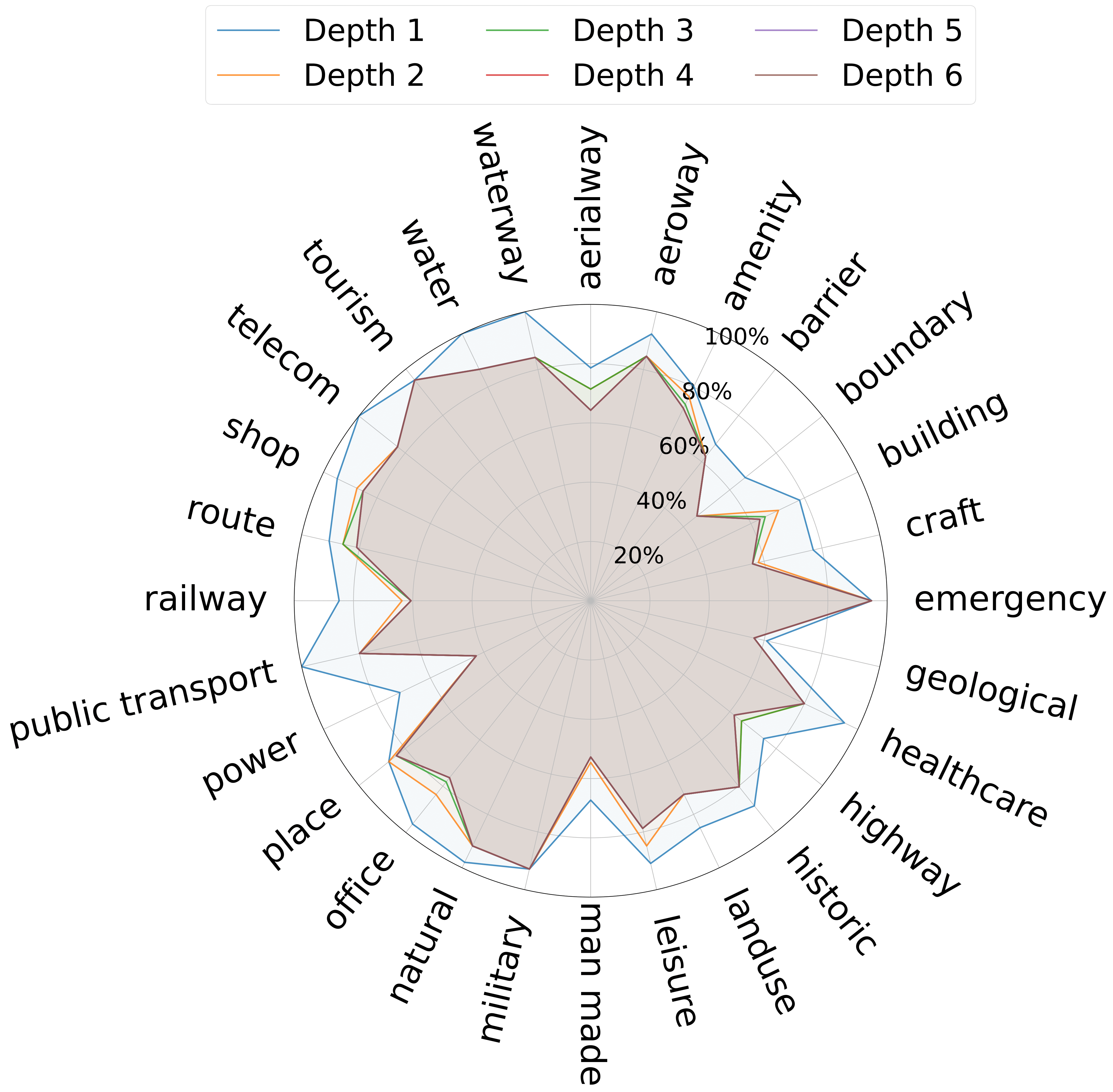}
    \caption{}
    \label{fig:k20_osm_d}
  \end{subfigure}
    \begin{subfigure}[]{0.5\textwidth}
    \vspace{0pt}
    \centering
    \includegraphics[width=\linewidth]{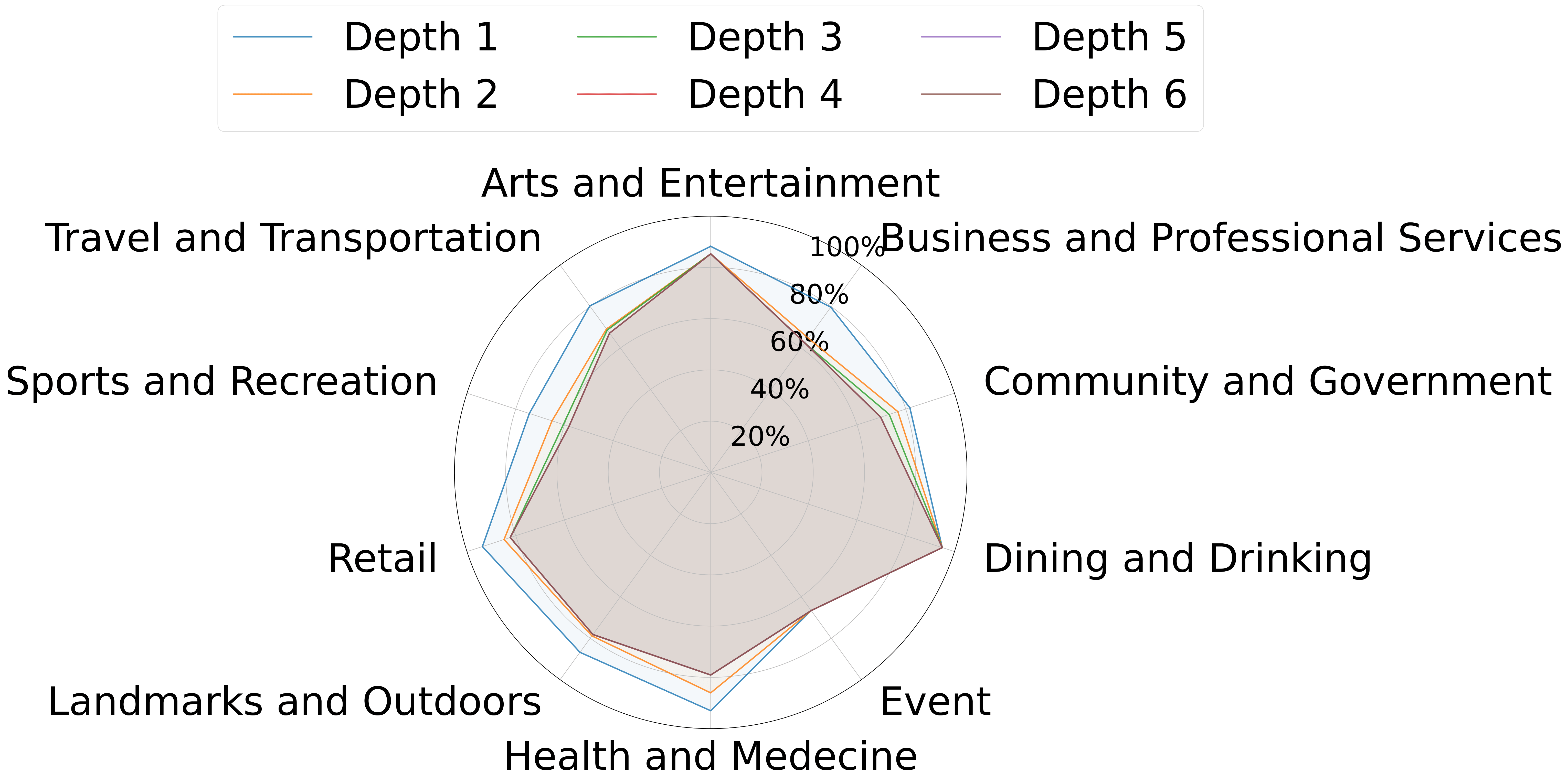}
    \caption{}
    \label{fig:k20_fs_d}
  \end{subfigure}
          \caption{Match percentages for OSM and FS categories at $k=20$ using the \texttt{Fallback - No example} prompt, broken down by hierarchy depth levels. On the left, the radar chart shows the match percentages across OSM tags, while on the right, the radar chart displays the corresponding match percentages for FS categories. }
  \label{fig:k_osm_fs_d}
\end{figure}

Among the prompting strategies, the first two consistently yield the highest recall across depths, confirming their robustness in guiding the refinement stage. Retaining $k=20$ ensures that the correct match is more likely to be present among the candidates while avoiding the performance degradation observed at larger shortlist sizes. Figure~\ref{fig:k_osm_fs} shows the match percentages categorized by OSM and FS categories for $k=20$ across the different prompting strategies, providing insight into which categories the models perform better or worse at matching. For instance, the models exhibit considerable differences in matching OSM tags such as \texttt{public transport}, \texttt{historic}, and \texttt{geological}. Regarding FS categories, the largest gaps occur in \texttt{Health and Medicine} and \texttt{Travel and Transportation}. The \texttt{Sports and Recreation} category, with fewer tags, has less impact but still contributes to the overall performance.\\
Based on these results, we select the \texttt{all-MiniLM-L6-v2} model (with FS category descriptions), refined with prompt \texttt{Fallback - No example} as the configuration achieving the best trade-off at $k=20$. This improves overall performance from 57.18\% to 72.78\%, corresponding to a relative improvement of 27.3\%. At the first and second hierarchy levels, the improvement goes from 78.59\% to 84.40\% and from 61.99\% to 74.94\%, with relative gains of 7.4\% and 20.9\%, respectively (Figure~\ref{fig:mdel_ref}). Figure~\ref{fig:k_osm_fs_d} further shows the match percentages for OSM and FS categories at $k=20$ using the \texttt{Fallback - No example} prompt, broken down by hierarchy depth levels, highlighting that several OSM categories are consistently well matched, at least at the top level, while others remain more challenging.\\
Before concluding, it is important to consider positional bias, a well-known issue of LLM-based judges \cite{shi2024judging}. In our implementation, this effect was partially mitigated by adopting the strategy proposed in \cite{zheng2024llmselector}, namely, avoiding the use of FS category IDs and instead presenting only the candidate options. To further evaluate its impact, we conducted a test by placing the correct solution in the first position and measured how this influenced the results (Table~\ref{tab:miniLM_firstpos}). Clearly, the model suffers from positional bias; indeed, performance improves, but only modestly, from +0.75\% at the first hierarchy level to +3.74\% at deeper levels.\\
Finally, by analyzing the categories that remain mismatched, it becomes clear that these are often borderline tags, such as those shown in Table~\ref{tab:borderline_examples}, while others are provided on GitHub as supplementary material. In such cases, the mismatch should not necessarily be attributed to a model error, but rather to the inherent ambiguity of tag interpretation. Therefore, even when the manually validated tag appears among the top-5, top-10, or top-20 candidates with the highest similarity scores (as in Table~\ref{tab:fs_topk_comparison} and Table~\ref{tab:fs_topk_comparison_large}), the model may still not select it as the most appropriate match. 

\begin{table*}[htbp]
  \centering
  \setlength{\tabcolsep}{4pt} 
  \renewcommand{\arraystretch}{0.9} 
  \caption{Model performance comparison across different depths with $k$ values.}
  \label{tab:model_performance_comparison_k}
  \begin{tabular}{l l S[table-format=2.2] S[table-format=2.2] S[table-format=2.2] S[table-format=2.2] S[table-format=2.2] S[table-format=2.2]}
    \toprule
    \textbf{\textbf{k}} & \textbf{Refinement} & \multicolumn{5}{c}{\textbf{Top-1 (\%)}} \\
    \cmidrule(lr){3-8}
    & & \textbf{Depth 1} & \textbf{Depth 2} & \textbf{Depth 3} & \textbf{Depth 4} & \textbf{Depth 5} & \textbf{Depth 6} \\
    \midrule
    5 & No fallback - No example & 84.23 & 73.94 & 72.03 & 71.70 & 71.70 & 71.70 \\
    & Fallback - No example & 84.15 & 73.53 & 71.37 & 70.95 & 70.95 & 70.95 \\
    & No fallback - Example & 83.15 & 72.70 & 70.62 & 69.88 & 69.88 & 69.88 \\
    & Fallback - Example & 84.15 & 71.87 & 69.96 & 69.29 & 69.29 & 69.29 \\
    \midrule
    {10} & No fallback - No example & 84.23 & 74.61 & 73.03 & 72.78 & 72.78 & 72.78 \\
    & Fallback - No example & 83.90 & 74.02 & 72.20 & 71.87 & 71.87 & 71.87 \\
    & No fallback - Example & 82.90 & 72.53 & 70.79 & 70.21 & 70.21 & 70.21 \\
    & Fallback - Example & 82.90 & 70.12 & 68.63 & 68.13 & 68.13 & 68.13 \\
    \midrule
    {20} & No fallback - No example & 83.57 & 74.27 & 72.12 & 71.54 & 71.54 & 71.54 \\
    & Fallback - No example & 84.40 & 74.94 & 73.03 & 72.28 & 72.28 & 72.28 \\
    & No fallback - Example & 82.74 & 72.70 & 70.12 & 69.29 & 69.29 & 69.29 \\
    & Fallback - Example & 82.90 & 70.46 & 68.05 & 67.14 & 67.14 & 67.14 \\
    \midrule
    {30} & No fallback - No example & 82.57 & 73.11 & 70.87 & 70.12 & 70.12 & 70.12 \\
    & Fallback - No example & 83.57 & 73.28 & 71.45 & 70.62 & 70.62 & 70.62 \\
    & No fallback - Example & 82.07 & 72.12 & 69.96 & 68.88 & 68.88 & 68.88 \\
    & Fallback - Example & 83.32 & 69.79 & 67.47 & 66.47 & 66.47 & 66.47 \\
    \midrule
    {40} & No fallback - No example & 82.16 & 72.28 & 70.21 & 69.54 & 69.54 & 69.54 \\
    & Fallback - No example & 83.24 & 73.28 & 71.70 & 70.87 & 70.87 & 70.87 \\
    & No fallback - Example & 83.15 & 72.61 & 70.37 & 69.29 & 69.29 & 69.29 \\
    & Fallback - Example & 82.32 & 69.96 & 67.88 & 66.80 & 66.80 & 66.80 \\
    \midrule
    {50} & No fallback - No example & 81.83 & 72.61 & 70.54 & 69.79 & 69.79 & 69.79 \\
    & Fallback - No example & 82.41 & 72.45 & 70.87 & 69.96 & 69.96 & 69.96 \\
    & No fallback - Example & 82.49 & 71.95 & 69.79 & 68.71 & 68.71 & 68.71 \\
    & Fallback - Example & 81.41 & 69.79 & 67.72 & 66.72 & 66.72 & 66.72 \\
    \bottomrule
  \end{tabular}
\end{table*}

\begin{table}[htbp]
  \centering
  \caption{\texttt{Fallback - No example} refinement performance (k=20) with the correct match placed in the first position.}
  \label{tab:miniLM_firstpos}
  \begin{tabular}{S[table-format=2.2] S[table-format=2.2] S[table-format=2.2] S[table-format=2.2] S[table-format=2.2] S[table-format=2.2]}
    \toprule
    \multicolumn{6}{c}{\textbf{Top-1 (\%)}} \\
    \cmidrule(lr){1-6}
    \textbf{Depth 1} & \textbf{Depth 2} & \textbf{Depth 3} & \textbf{Depth 4} & \textbf{Depth 5} & \textbf{Depth 6} \\
    \midrule
    85.15 & 77.84 & 76.51 & 76.02 & 76.02 & 76.02 \\
    \bottomrule
  \end{tabular}
\end{table}

\newcolumntype{Y}{>{\raggedright\arraybackslash}X} 

\begin{table*}[htbp]
  \centering
  
  \setlength{\tabcolsep}{4pt}   
  \renewcommand{\arraystretch}{0.95} 
  \caption{Examples of borderline OSM tags and their FS mappings.}
  \label{tab:borderline_examples}
  \begin{tabularx}{\textwidth}{l l Y Y}
    \toprule
    \textbf{OSM tag}  & \textbf{OSM Depth 1} & \textbf{FS manual match} & \textbf{FS model match} \\
    \midrule
    dive centre & amenity & sports recreation $\rightarrow$ water sports $\rightarrow$ scuba diving instructor & landmarks outdoors $\rightarrow$ dive spot\\
    riding hall & building & sports recreation $\rightarrow$ equestrian facility & landmarks outdoors $\rightarrow$ stable\\
    leather     & craft    & business professional services $\rightarrow$ leather supplier & retail $\rightarrow$ leather goods store\\
    engraver    & craft    & business professional services $\rightarrow$ creative service & retail $\rightarrow$ arts crafts store\\
    jeweller    & craft    & business professional services $\rightarrow$ creative service & retail $\rightarrow$ fashion retail $\rightarrow$ jewelry store\\
    optician    & craft    & health medicine $\rightarrow$ optometrist & retail $\rightarrow$ eyecare store\\
    bag         & shop     & retail $\rightarrow$ luggage store & retail $\rightarrow$ fashion retail $\rightarrow$ fashion accessories store \\
    music       & shop     & retail $\rightarrow$ record store & retail $\rightarrow$ music store \\
    sewing      & shop     & retail $\rightarrow$ knitting store & retail $\rightarrow$ textiles store \\
    wool        & shop     & retail $\rightarrow$ knitting store & retail $\rightarrow$ textiles store \\
    \bottomrule
  \end{tabularx}
\end{table*}
\subsection*{Types of mismatches}
In this subsection, we present a qualitative examination of the mismatches and group them into a set of main categories, each illustrated with representative examples.
\subsubsection*{Lexical-description mismatch}
A source of mismatch arises when the literal name of an OSM tag and its official description suggest different interpretations.  
In some cases, the model prioritizes the lexical form of the tag, while the benchmark follows the meaning clarified in the description.  \\
For example, the OSM tag \texttt{farm} is mapped by the model to \texttt{landmarks and outdoors > farm}, reflecting a direct lexical association.  
However, according to the OSM documentation, this tag specifically refers to a \textit{farmhouse}, i.e., a residential building located on a farm.  
In the benchmark, this led to the correspondence with \texttt{community and government > residential building}, judged more appropriate given the descriptive context.  
\subsubsection*{Lexical bias in category assignment}
A second type of mismatch occurs when the model prioritizes the literal form of a tag over its broader semantic scope.  
In these cases, the lexical match drives the alignment toward categories that share the same wording, even when the actual domain of the tag points elsewhere.  \\
For example, OSM tags under Depth~1 \texttt{geological} are consistently mapped by the model to \texttt{business professional services > geological service}, reflecting a direct lexical association.  
In the manual benchmark, however, these tags were linked to \texttt{landmarks and outdoors}, since tags such as \texttt{rock glacier}, \texttt{meteor crater}, or \texttt{moraine} describe natural landforms rather than professional services.  \\
This illustrates how lexical bias can overshadow semantic context: the model aligns to the closest lexical match, whereas the benchmark emphasizes the actual domain of use.  
\subsubsection*{Multiple valid correspondences}
In some situations, more than one FS category can be considered plausible, making a single “correct” mapping debatable.  
For instance, the OSM tag \texttt{music}, described as \textit{“Shop focused on selling recorded music (vinyl/CDs/...)”}, could validly match both \texttt{retail > record store} and \texttt{retail > music store}. 
A similar case occurs with OSM tags such as \texttt{wool} or \texttt{sewing}, which may correspond either to \texttt{knitting store} or to \texttt{textiles store}.  
Since these two FS categories are at the same hierarchical level, the ambiguity does not reflect a difference in granularity but rather the high degree of detail in the FS taxonomy, where multiple equally valid categories may coexist for the same OSM tag.  
\subsubsection*{Granularity mismatches}
Finally, some discrepancies arise from differences in the level of granularity between OSM and FS. In such cases, it may be reasonable to accept matches at a broader level of the hierarchy, or alternatively to allow multiple subcategories. However, this remains a limitation of relying exclusively on category names and automatically generated descriptions.
\subsection*{Interpretation and implications}
In this work, we introduced a manually validated ground truth for mapping OSM and FS categories, which constitutes a valuable contribution both as a benchmark for evaluating alignment algorithms and as a resource for developing more reliable methodologies for integrating heterogeneous geospatial sources. \\
The FS taxonomy was chosen as the reference structure because it is hierarchical and well-curated, allowing analyses to be conducted at different levels of depth depending on the research focus.  
This hierarchical dimension is essential for combining the richness of community-driven OSM contributions with the consistency required for reproducible multi-scale analyses.
Building on this foundation, we proposed an LLM-based refinement pipeline that combines embedding retrieval with generative re-ranking. This design choice is particularly relevant because OSM is continuously evolving: new tags are added, descriptions are refined, and usage patterns change over time.  
The proposed pipeline therefore provides not only a reproducible benchmark at a given point in time, but also a scalable and adaptable methodology that can update mappings as the underlying taxonomy evolves.
Our results show that this approach substantially improves alignment accuracy compared to baseline models without the refinement stage, enabling the mapping of around 85\% of OSM tags at the first hierarchy level and over 72\% at the deepest levels, with relative gains of up to 27\%.\\
A qualitative analysis of residual mismatches showed that most discrepancies are not systematic errors but fall into a limited set of recurrent patterns. We identified four main types of mismatches between manual and automatic alignment, which highlight the intrinsic challenges of reconciling heterogeneous taxonomies. These mismatches often stem from lexical ambiguities, conflicts between tag names and their intended meaning, or strong differences in granularity between the two datasets, rather than from genuine model failures.
This reinforces the idea that the proposed approach provides a solid basis for the integration of geospatial datasets, while also pointing to the need for solutions that improve semantic clarity.  \\
A possible way forward would be the availability of official documentation for FS categories, similar to the comprehensive resources already provided for OSM. Carefully curated and openly shared descriptions would allow for better disambiguation of concepts and ensure more robust matches. These descriptions could also complement or replace those generated with ChatGPT, which, although useful for bootstrapping, remain inevitably weaker from a semantic perspective. Human-in-the-loop validation can further enhance reliability by resolving borderline cases where multiple mappings may be equally plausible.  \\
Another relevant limitation concerns the \textit{positional bias} of LLMs, namely their tendency to favor items presented earlier in a list \cite{shi2024judging, guo2024biasllms}. This effect was partially mitigated in our implementation by following the suggestion of \cite{zheng2024llmselector}, namely, avoiding the use of FS category IDs and instead presenting only the candidate options. This design choice also reduces the typical error of LLMs that occurs when they develop a correct reasoning process but then fail to select the correct option, ultimately providing an alternative instead \cite{molfese2025right}. Further improvements could come from implementing prompting strategies that explicitly shuffle the order of candidates or from evaluation schemes that normalize for positional effects \cite{li2023generative}. However, while positional bias remains a well-known challenge in the design of robust refinement pipelines, in our case it did not have a substantial impact on the final results.\\
Beyond technical accuracy, what emerges from this work is a broader perspective on how collective and curated knowledge can complement each other in shaping the way we represent cities. OSM contributes richness and granularity rooted in local participation, with particular strength in describing natural features and mobility infrastructures, while FS provides consistency and hierarchy, especially in representing urban services and commercial activities. Bringing these two paradigms together opens the way to more comprehensive urban datasets, capable of supporting mobility research, urban planning, and smart city applications with greater reliability and consistency.  \\
Overall, the proposed methodology demonstrates the potential of LLMs for advancing data integration in urban analytics, enabling richer and more interoperable datasets. Future work should focus on extending the benchmark, incorporating more contextual information such as full category hierarchies, experimenting with stronger models and alternative prompting strategies, and including human-in-the-loop validation. Looking ahead, extending the framework to other platforms, languages, and the integration of dynamic updates will be crucial. Yet the central message remains: unifying taxonomies is not only a matter of technical interoperability, but also a step towards richer, more inclusive ways of seeing, analyzing, and comprehending the complexity of urban life.

\section*{Data Availability}

The raw datasets used in this study are openly available from their original sources. 
OpenStreetMap (OSM) data are publicly accessible under the Open Database License (ODbL) at \cite{openstreetmap}.  
The Foursquare Open Source Places dataset is available under the Apache 2.0 License at \cite{foursquare_open}. \\ 
Processed datasets generated in this study, including the cleaned OSM and FS taxonomies, the manually curated benchmark mapping, enriched FS category descriptions, and all intermediate alignment results, are openly available in the project repository \cite{osmfs_benchmark_2025}.  
The repository contains cleaned category tables, the oracle benchmark, embedding predictions, refinement outputs, and all materials necessary to reproduce the analyses.

\section*{Code availability}
All code developed for this study, including the scraping pipeline for OSM documentation and the taxonomy alignment framework, is openly available in the project repository \cite{osmfs_benchmark_2025}. The repository provides preprocessing scripts, benchmark mappings, and the full implementation of the alignment framework described in this manuscript. The code is publicly accessible at the link provided in the reference and is permanently archived and linked to this article.

\section*{Funding}
E.A. and M.N. acknowledges the financial support received from the TALEA Green Cells Leading the Green Transition (EUI02-064) project. S.B., S.C., R.G., B.L. and M.N. acknowledge the support of the PNRR ICSC National Research Centre for High Performance Computing, Big Data and Quantum Computing (CN00000013), under the NRRP MUR program funded by the NextGenerationEU.

\section*{Author Contributions}
L. Soulas conducted all analyses, including preprocessing scripts, benchmark mappings, and the alignment framework described in this paper.  
L. Lucchini contributed to the development of the web scraping code.  
M. Napolitano, as an OpenStreetMap expert, advised on methodological choices for setting up the mapping.  
E. Andreotti designed the study, wrote the manuscript, and revised the analyses.  
S. Bontorin, S. Centellegher, B. Lepri and R. Gallotti provided supervision and useful discussions.  
All authors discussed the results and approved the final version of the manuscript.

\bibliographystyle{unsrtnat}
\bibliography{embeddedbib}

\begin{thebibliography}{44}
\providecommand{\natexlab}[1]{#1}
\providecommand{\url}[1]{\texttt{#1}}
\expandafter\ifx\csname urlstyle\endcsname\relax
  \providecommand{\doi}[1]{doi: #1}\else
  \providecommand{\doi}{doi: \begingroup \urlstyle{rm}\Url}\fi

\bibitem[Lucchini et~al.(2021)Lucchini, Centellegher, Pappalardo, Gallotti, Privitera, Lepri, and De~Nadai]{lucchini2021living}
Lorenzo Lucchini, Simone Centellegher, Luca Pappalardo, Riccardo Gallotti, Filippo Privitera, Bruno Lepri, and Marco De~Nadai.
\newblock Living in a pandemic: changes in mobility routines, social activity and adherence to covid-19 protective measures.
\newblock \emph{Scientific Reports}, 11\penalty0 (1):\penalty0 24452, December 2021.
\newblock \doi{10.1038/s41598-021-04139-1}.

\bibitem[Moro et~al.(2021)Moro, Calacci, Dong, and Pentland]{moro2021mobility}
Esteban Moro, Dan Calacci, Xiaowen Dong, and Alex Pentland.
\newblock Mobility patterns are associated with experienced income segregation in large us cities.
\newblock \emph{Nature communications}, 12\penalty0 (1):\penalty0 4633, 2021.

\bibitem[Fraser et~al.(2024)Fraser, Yabe, Aldrich, and Moro]{fraser2024great}
Timothy Fraser, Takahiro Yabe, Daniel~P Aldrich, and Esteban Moro.
\newblock The great equalizer? mixed effects of social infrastructure on diverse encounters in cities.
\newblock \emph{Computers, Environment and Urban Systems}, 113:\penalty0 102173, 2024.

\bibitem[Centellegher et~al.(2025)Centellegher, De~Nadai, Tonin, Lepri, and Lucchini]{centellegher2025job}
Simone Centellegher, Marco De~Nadai, Marco Tonin, Bruno Lepri, and Lorenzo Lucchini.
\newblock Job loss disrupts individuals’ mobility and their exploratory patterns.
\newblock \emph{iScience}, 2025.

\bibitem[Bontorin et~al.(2025)Bontorin, Centellegher, Gallotti, Pappalardo, Lepri, and Luca]{bontorin2025mixing}
Sebastiano Bontorin, Simone Centellegher, Riccardo Gallotti, Luca Pappalardo, Bruno Lepri, and Massimiliano Luca.
\newblock Mixing individual and collective behaviors to predict out-of-routine mobility.
\newblock \emph{Proceedings of the National Academy of Sciences}, 122\penalty0 (17):\penalty0 e2414848122, 2025.

\bibitem[Li et~al.(2023{\natexlab{a}})Li, Zhang, Zhao, Liu, Zhang, and Ratti]{Li2023StreetGreenery}
X.~Li, C.~Zhang, Q.~Zhao, L.~Liu, Y.~Zhang, and C.~Ratti.
\newblock A global high-resolution dataset of street-level greenery and walkability indicators with deep learning.
\newblock \emph{Scientific Data}, 10\penalty0 (1):\penalty0 712, 2023{\natexlab{a}}.
\newblock \doi{10.1038/s41597-023-02576-3}.

\bibitem[Andreotti et~al.(2025)Andreotti, Usmani, and Napolitano]{andreotti2025city}
Eleonora Andreotti, Munazza Usmani, and Maurizio Napolitano.
\newblock The city as a complex system: Balancing identity and infrastructure needs.
\newblock Manuscript in preparation, August 2025.

\bibitem[{Google LLC}(2025)]{googleplaces}
{Google LLC}.
\newblock Google places api documentation.
\newblock \url{https://developers.google.com/maps/documentation/places/web-service/overview}, 2025.
\newblock Accessed: November 2025.

\bibitem[{SafeGraph Inc.}(2025)]{safegraph}
{SafeGraph Inc.}
\newblock Safegraph places and patterns datasets.
\newblock \url{https://www.safegraph.com/}, 2025.
\newblock Accessed: November 2025.

\bibitem[Haklay and Weber(2008)]{haklay2008osm}
Mordechai Haklay and Patrick Weber.
\newblock Openstreetmap: User-generated street maps.
\newblock \emph{IEEE Pervasive Computing}, 7\penalty0 (4):\penalty0 12--18, 2008.
\newblock \doi{10.1109/MPRV.2008.80}.

\bibitem[fou()]{foursquare_open}
Foursquare open source places - os places dataset.
\newblock \url{https://opensource.foursquare.com/os-places/}.
\newblock Accessed: 2025-11-03.

\bibitem[{Foursquare Labs Inc.}(2025)]{foursquare}
{Foursquare Labs Inc.}
\newblock Foursquare places api, 2025.
\newblock URL \url{https://developer.foursquare.com/places}.
\newblock Accessed: 2025-08-17.

\bibitem[{Overture Maps Foundation}(2025)]{overture}
{Overture Maps Foundation}.
\newblock Overture maps foundation -- places dataset.
\newblock \url{https://docs.overturemaps.org/guides/places/}, 2025.
\newblock Accessed: November 2025.

\bibitem[Priem et~al.(2022)Priem, Piwowar, and Orr]{priem2022openalex}
Jason Priem, Heather Piwowar, and Richard Orr.
\newblock Openalex: A fully-open index of scholarly works, authors, venues, institutions, and concepts.
\newblock \emph{arXiv preprint arXiv:2205.01833}, 2022.
\newblock Accessed via the OpenAlex documentation.

\bibitem[{OpenAlex}(2025)]{openalex2025}
{OpenAlex}.
\newblock Openalex database.
\newblock \url{https://openalex.org/}, 2025.
\newblock Accessed: October 20, 2025.

\bibitem[Zhu(2012)]{zhu2012survey}
Junwu Zhu.
\newblock Survey on ontology mapping.
\newblock \emph{Physics Procedia}, 24\penalty0 (Part C):\penalty0 1857--1862, 2012.
\newblock ISSN 1875-3892.
\newblock \doi{10.1016/j.phpro.2012.02.274}.

\bibitem[Shbita and Knoblock(2024)]{shbita2024osm}
A.~Shbita and C.~A. Knoblock.
\newblock Automatically constructing geospatial feature taxonomies from openstreetmap data.
\newblock In \emph{2024 IEEE International Conference on Semantic Computing (ICSC)}, pages 149--156. IEEE, 2024.
\newblock \doi{10.1109/ICSC59869.2024.00035}.

\bibitem[Zhang and Pfoser(2019)]{zhang2019osmfs}
L.~Zhang and D.~Pfoser.
\newblock Using openstreetmap point-of-interest data to model urban change -- a feasibility study.
\newblock \emph{International Journal of Geo-Information}, 8\penalty0 (1):\penalty0 1--19, 2019.
\newblock \doi{10.3390/ijgi8010037}.

\bibitem[Novack et~al.(2018)Novack, Peters, and Zipf]{novack2018graph}
T.~Novack, R.~Peters, and A.~Zipf.
\newblock Graph-based matching of points-of-interest from collaborative geo-datasets.
\newblock In \emph{Proceedings of the 21st AGILE International Conference on Geographic Information Science}, pages 1--17. Springer, 2018.
\newblock \doi{10.1007/978-3-319-78208-9_17}.

\bibitem[Steiner et~al.(2025)Steiner, Peeters, and Bizer]{11107461}
Aaron Steiner, Ralph Peeters, and Christian Bizer.
\newblock Fine-tuning large language models for entity matching.
\newblock In \emph{2025 IEEE 41st International Conference on Data Engineering Workshops (ICDEW)}, pages 9--17, 2025.
\newblock \doi{10.1109/ICDEW67478.2025.00006}.

\bibitem[Hertling and Paulheim(2023)]{10.1145/3587259.3627571}
Sven Hertling and Heiko Paulheim.
\newblock Olala: Ontology matching with large language models.
\newblock In \emph{Proceedings of the 12th Knowledge Capture Conference 2023}, K-CAP '23, page 131–139, New York, NY, USA, 2023. Association for Computing Machinery.
\newblock ISBN 9798400701412.
\newblock \doi{10.1145/3587259.3627571}.
\newblock URL \url{https://doi.org/10.1145/3587259.3627571}.

\bibitem[Taboada et~al.(2025)Taboada, Martinez, Arideh, and Mosquera]{TABOADA2025103254}
Maria Taboada, Diego Martinez, Mohammed Arideh, and Rosa Mosquera.
\newblock Ontology matching with large language models and prioritized depth-first search.
\newblock \emph{Information Fusion}, 123:\penalty0 103254, 2025.
\newblock ISSN 1566-2535.
\newblock \doi{https://doi.org/10.1016/j.inffus.2025.103254}.
\newblock URL \url{https://www.sciencedirect.com/science/article/pii/S1566253525003276}.

\bibitem[Le et~al.(2025)Le, Nguyen, Ha, Chatzinotas, Jouvet, and Noumeir]{LeHaNguyenEtAl2025}
Thanh-Dung Le, Ti~Ti Nguyen, Vu~Nguyen Ha, Symeon Chatzinotas, Philippe Jouvet, and Rita Noumeir.
\newblock The impact of lora adapters on llms for clinical text classification under computational and data constraints.
\newblock \emph{IEEE Access}, 13:\penalty0 109365--109377, 2025.
\newblock \doi{10.1109/ACCESS.2025.3582037}.
\newblock Received 27 May 2025; accepted 16 June 2025; published 24 June 2025; current version 1 July 2025.

\bibitem[OpenAI(2025{\natexlab{a}})]{openai_gpt4omini}
OpenAI.
\newblock Gpt-4o-mini via openai api [large language model].
\newblock \url{https://platform.openai.com}, 2025{\natexlab{a}}.
\newblock Accessed: 2025-09-15.

\bibitem[OpenAI(2025{\natexlab{b}})]{openai_gpt5mini}
OpenAI.
\newblock Gpt-5-mini via openai api [large language model].
\newblock \url{https://platform.openai.com}, 2025{\natexlab{b}}.
\newblock Accessed: 2025-09-15.

\bibitem[OSM(2025)]{OSM_DeprecatedFeatures_2025}
Deprecated features.
\newblock OpenStreetMap Wiki, 2025.
\newblock URL \url{{https://wiki.openstreetmap.org/wiki/Deprecated_features}}.
\newblock Online; accessed 10 September 2025.

\bibitem[Soulas(2025)]{osmfs_benchmark_2025}
Lilou Soulas.
\newblock Benchmark mapping of osm and fs categories.
\newblock \url{https://github.com/LilouSoulas/Mapping-of-OSM-and-FS-categories}, 2025.
\newblock Accessed: 2025-09-10.

\bibitem[Reimers and Gurevych(2019)]{reimers2019sentence}
Nils Reimers and Iryna Gurevych.
\newblock Sentence-bert: Sentence embeddings using siamese bert-networks.
\newblock In \emph{Proceedings of the 2019 Conference on Empirical Methods in Natural Language Processing}, pages 3982--3992. Association for Computational Linguistics, 2019.
\newblock \doi{10.18653/v1/D19-1410}.

\bibitem[Wang et~al.(2020)Wang, Wei, Dong, Bao, Yang, and Zhou]{wang2020minilm}
Wenhui Wang, Furu Wei, Li~Dong, Hangbo Bao, Nan Yang, and Ming Zhou.
\newblock Minilm: Deep self-attention distillation for task-agnostic compression of pre-trained transformers.
\newblock In \emph{Advances in Neural Information Processing Systems}, volume~33, pages 5776--5788, 2020.

\bibitem[Sanh et~al.(2019)Sanh, Debut, Chaumond, and Wolf]{sanh2019distilbert}
Victor Sanh, Lysandre Debut, Julien Chaumond, and Thomas Wolf.
\newblock Distilbert, a distilled version of bert: smaller, faster, cheaper and lighter.
\newblock In \emph{Proceedings of the 5th Workshop on Energy Efficient Machine Learning and Cognitive Computing (NeurIPS)}, 2019.

\bibitem[Liu et~al.(2019)Liu, Ott, Goyal, and et~al.]{liu2019roberta}
Yinhan Liu, Myle Ott, Naman Goyal, and et~al.
\newblock Roberta: A robustly optimized bert pretraining approach.
\newblock In \emph{arXiv:1907.11692}, 2019.

\bibitem[Jiao et~al.(2020)Jiao, Yin, Shang, Jiang, Chen, Li, Wang, and Gong]{jiao2020tinybert}
Xiaoqi Jiao, Yichun Yin, Lifeng Shang, Xin Jiang, Xiao Chen, Linlin Li, Fang Wang, and Furu Gong.
\newblock Tinybert: Distilling bert for natural language understanding.
\newblock In \emph{Findings of the Association for Computational Linguistics: EMNLP 2020}, pages 4163--4174, 2020.

\bibitem[Song et~al.(2020)Song, Tan, Qin, Lu, and Liu]{song2020mpnet}
Kaitao Song, Xu~Tan, Tao Qin, Jianfeng Lu, and Tie-Yan Liu.
\newblock Mpnet: Masked and permuted pre-training for language understanding.
\newblock In \emph{Advances in Neural Information Processing Systems}, volume~33, pages 16857--16867, 2020.

\bibitem[OpenAI(2025{\natexlab{c}})]{openai_chatgpt5}
OpenAI.
\newblock Chatgpt (gpt-5) [large language model].
\newblock \url{https://chat.openai.com}, 2025{\natexlab{c}}.
\newblock Accessed: 2025-09-15.

\bibitem[Mardia et~al.(1979)Mardia, Kent, and Bibby]{mardia1979multivariate}
Kanti~V. Mardia, John~T. Kent, and John~M. Bibby.
\newblock \emph{Multivariate Analysis}.
\newblock Academic Press, London, 1979.
\newblock ISBN 9780124712522.

\bibitem[Sokolova and Lapalme(2006)]{sokolova2006systematic}
M.~Sokolova and G.~Lapalme.
\newblock A systematic analysis of performance measures for classification tasks.
\newblock \emph{Information Processing \& Management}, 42\penalty0 (1):\penalty0 247--265, 2006.

\bibitem[Fawcett(2006)]{fawcett2006roc}
Tom Fawcett.
\newblock An introduction to roc analysis.
\newblock \emph{Pattern Recognition Letters}, 27\penalty0 (8):\penalty0 861--874, 2006.

\bibitem[{Sentence Transformers}(2025)]{sbert_pretrained_models}
{Sentence Transformers}.
\newblock Pretrained models — sentence transformers.
\newblock Online documentation, 2025.
\newblock URL \url{\url{https://www.sbert.net/docs/sentence\_transformer/pretrained\_models.html}}.
\newblock Accessed: 2025-09-16.

\bibitem[Shi et~al.(2025)Shi, Ma, Liang, Ma, and Vosoughi]{shi2024judging}
Lin Shi, Chiyu Ma, Wenhua Liang, Weicheng Ma, and Soroush Vosoughi.
\newblock Judging the judges: A systematic study of position bias in llm-as-a-judge.
\newblock \emph{arXiv preprint arXiv:2406.07791}, 2025.
\newblock URL \url{https://arxiv.org/abs/2406.07791}.

\bibitem[Zheng et~al.(2024)Zheng, Zhou, Meng, Zhou, and Huang]{zheng2024llmselector}
Chujie Zheng, Hao Zhou, Fandong Meng, Jie Zhou, and Minlie Huang.
\newblock Large language models are not robust multiple choice selectors.
\newblock In \emph{Proceedings of the International Conference on Learning Representations (ICLR)}, 2024.

\bibitem[Guo et~al.(2024)Guo, Guo, Su, Yang, Zhu, Li, Qiu, and Liu]{guo2024biasllms}
Yufei Guo, Muzhe Guo, Juntao Su, Zhou Yang, Mengqiu Zhu, Hongfei Li, Mengyang Qiu, and Shuo~Shuo Liu.
\newblock Bias in large language models: Origin, evaluation, and mitigation.
\newblock \emph{arXiv preprint arXiv:2411.10915}, 2024.
\newblock \doi{10.48550/arXiv.2411.10915}.
\newblock URL \url{https://arxiv.org/abs/2411.10915}.

\bibitem[Molfese et~al.(2025)Molfese, Moroni, Gioffré, Scirè, Conia, and Navigli]{molfese2025right}
Francesco~Maria Molfese, Luca Moroni, Luca Gioffré, Alessandro Scirè, Simone Conia, and Roberto Navigli.
\newblock Right answer, wrong score: Uncovering the inconsistencies of llm evaluation in multiple-choice question answering, 2025.
\newblock URL \url{https://arxiv.org/abs/2503.14996}.

\bibitem[Li et~al.(2023{\natexlab{b}})Li, Sun, Yuan, Fan, Zhao, and Liu]{li2023generative}
Junlong Li, Shichao Sun, Weizhe Yuan, Run-Ze Fan, Hai Zhao, and Pengfei Liu.
\newblock Generative judge for evaluating alignment.
\newblock \emph{arXiv preprint arXiv:2310.05470}, 2023{\natexlab{b}}.
\newblock URL \url{https://arxiv.org/abs/2310.05470}.

\bibitem[{OpenStreetMap contributors}(2025)]{openstreetmap}
{OpenStreetMap contributors}.
\newblock Openstreetmap, 2025.
\newblock URL \url{https://www.openstreetmap.org}.
\newblock Accessed: 2025-08-17.

\end{thebibliography}
\newpage
\section{Main-category manual matches}\label{asec:tabmainmatch}

\begin{table}[!ht]
\caption{Mappings for OSM Depth 1 = aerialway}\label{taba:13}
\centering
\begin{tabular}{l|l}
\toprule
\textbf{OSM tag} & \textbf{FS manual match} \\
\midrule
zip line & sports recreation \\
\bottomrule
\end{tabular}
\end{table}

\begin{table}[!ht]
\caption{Mappings for OSM Depth 1 = aeroway}\label{taba:14}
\centering
\begin{tabular}{l|l}
\toprule
\textbf{OSM tag} & \textbf{FS manual match} \\
\midrule
windsock & landmarks outdoors \\
spaceport & travel transportation \\
\bottomrule
\end{tabular}
\end{table}

\begin{table}[!ht]
\caption{Mappings for OSM Depth 1 = amenity}\label{taba:15}
\centering
\begin{tabular}{l|l}
\toprule
\textbf{OSM tag} & \textbf{FS manual match} \\
\midrule
brothel & arts entertainment \\
social centre & arts entertainment \\
swingerclub & arts entertainment \\
bicycle repair station & business professional services \\
bicycle wash & business professional services \\
dressing room & business professional services \\
kitchen & business professional services \\
baby hatch & community government \\
give box & community government \\
shelter & community government \\
shower & community government \\
bbq & landmarks outdoors \\
bench & landmarks outdoors \\
kneipp water cure & landmarks outdoors \\
lounger & landmarks outdoors \\
watering place & landmarks outdoors \\
animal training & sports recreation \\
boat sharing & travel transportation \\
car sharing & travel transportation \\
weighbridge & travel transportation \\
\bottomrule
\end{tabular}
\end{table}

\begin{table}[!ht]
\caption{Mappings for OSM Depth 1 = barrier}\label{taba:16}
\centering
\begin{tabular}{l|l}
\toprule
\textbf{OSM tag} & \textbf{FS manual match} \\
\midrule
handrail & landmarks outdoors \\
motorcycle barrier & landmarks outdoors \\
block & travel transportation \\
bollard & travel transportation \\
bump gate & travel transportation \\
bus trap & travel transportation \\
cable barrier & travel transportation \\
cattle grid & travel transportation \\
cycle barrier & travel transportation \\
ditch & travel transportation \\
entrance & travel transportation \\
fullheight turnstile & travel transportation \\
gate & travel transportation \\
guard rail & travel transportation \\
height restrictor & travel transportation \\
horse stile & travel transportation \\
jersey barrier & travel transportation \\
kissing gate & travel transportation \\
lift gate & travel transportation \\
log & travel transportation \\
rope & travel transportation \\
spikes & travel transportation \\
stile & travel transportation \\
sump buster & travel transportation \\
swing gate & travel transportation \\
turnstile & travel transportation \\
\bottomrule
\end{tabular}
\end{table}

\begin{table}[!ht]
\caption{Mappings for OSM Depth 1 = boundary}\label{taba:17}
\centering
\begin{tabular}{l|l}
\toprule
\textbf{OSM tag} & \textbf{FS manual match} \\
\midrule
health & health medicine \\
disputed & landmarks outdoors \\
hazard & landmarks outdoors \\
low emission zone & landmarks outdoors \\
limited traffic zone & travel transportation \\
\bottomrule
\end{tabular}
\end{table}

\begin{table}[!ht]
\caption{Mappings for OSM Depth 1 = building}\label{taba:18}
\centering
\begin{tabular}{l|l}
\toprule
\textbf{OSM tag} & \textbf{FS manual match} \\
\midrule
service & business professional services \\
civic & community government \\
conservatory & community government \\
gatehouse & community government \\
outbuilding & community government \\
roof & community government \\
tower & community government \\
quonset hut & landmarks outdoors \\
ruins & landmarks outdoors \\
ship & landmarks outdoors \\
trullo & landmarks outdoors \\
water tower & landmarks outdoors \\
commercial & retail \\
kiosk & retail \\
grandstand & sports recreation \\
\bottomrule
\end{tabular}
\end{table}

\begin{table}[!ht]
\caption{Mappings for OSM Depth 1 = craft}\label{taba:19}
\centering
\begin{tabular}{l|l}
\toprule
\textbf{OSM tag} & \textbf{FS manual match} \\
\midrule
builder & business professional services \\
cooper & business professional services \\
elevator & business professional services \\
oil mill & business professional services \\
paver & business professional services \\
rigger & business professional services \\
signmaker & business professional services \\
stand builder & business professional services \\
water well drilling & business professional services \\
dental technician & health medicine \\
boatbuilder & travel transportation \\
\bottomrule
\end{tabular}
\end{table}

\begin{table}[!ht]
\caption{Mappings for OSM Depth 1 = geological}\label{taba:20}
\centering
\begin{tabular}{l|l}
\toprule
\textbf{OSM tag} & \textbf{FS manual match} \\
\midrule
columnar jointing & landmarks outdoors \\
cone & landmarks outdoors \\
dyke & landmarks outdoors \\
fault & landmarks outdoors \\
fold & landmarks outdoors \\
giants kettle & landmarks outdoors \\
hoodoo & landmarks outdoors \\
limestone pavement & landmarks outdoors \\
meteor crater & landmarks outdoors \\
monocline & landmarks outdoors \\
moraine & landmarks outdoors \\
outcrop & landmarks outdoors \\
rock glacier & landmarks outdoors \\
sinkhole & landmarks outdoors \\
tor & landmarks outdoors \\
unconformity & landmarks outdoors \\
\bottomrule
\end{tabular}
\end{table}

\begin{table}[!ht]
\caption{Mappings for OSM Depth 1 = highway}\label{taba:21}
\centering
\begin{tabular}{l|l}
\toprule
\textbf{OSM tag} & \textbf{FS manual match} \\
\midrule
corridor & travel transportation \\
elevator & travel transportation \\
ladder & travel transportation \\
passing place & travel transportation \\
path & travel transportation \\
steps & travel transportation \\
traffic signals & travel transportation \\
\bottomrule
\end{tabular}
\end{table}

\begin{table}[!ht]
\caption{Mappings for OSM Depth 1 = historic}\label{taba:22}
\centering
\begin{tabular}{l|l}
\toprule
\textbf{OSM tag} & \textbf{FS manual match} \\
\midrule
shieling & landmarks outdoors \\
\bottomrule
\end{tabular}
\end{table}

\begin{table}[!t]
\caption{Mappings for OSM Depth 1 = landuse}\label{taba:23}
\centering
\begin{tabular}{l|l}
\toprule
\textbf{OSM tag} & \textbf{FS manual match} \\
\midrule
commercial & business professional services \\
construction & community government \\
institutional & community government \\
grass & landmarks outdoors \\
greenhouse horticulture & landmarks outdoors \\
salt pond & landmarks outdoors \\
depot & travel transportation \\
\bottomrule
\end{tabular}
\end{table}

\begin{table}[!t]
\caption{Mappings for OSM Depth 1 = leisure}\label{taba:24}
\centering
\begin{tabular}{l|l}
\toprule
\textbf{OSM tag} & \textbf{FS manual match} \\
\midrule
bird hide & landmarks outdoors \\
common & sports recreation \\
\bottomrule
\end{tabular}
\end{table}

\begin{table}[!t]
\caption{Mappings for OSM Depth 1 = man made}\label{taba:25}
\centering
\begin{tabular}{l|l}
\toprule
\textbf{OSM tag} & \textbf{FS manual match} \\
\midrule
monitoring station & business professional services \\
pipeline & business professional services \\
chimney & community government \\
guard stone & community government \\
kiln & community government \\
mast & community government \\
tower & community government \\
bunker silo & landmarks outdoors \\
embankment & landmarks outdoors \\
watermill & landmarks outdoors \\
wildlife crossing & landmarks outdoors \\
\bottomrule
\end{tabular}
\end{table}

\begin{table}[!t]
\caption{Mappings for OSM Depth 1 = natural}\label{taba:26}
\centering
\begin{tabular}{l|l}
\toprule
\textbf{OSM tag} & \textbf{FS manual match} \\
\midrule
arch & landmarks outdoors \\
arete & landmarks outdoors \\
bare rock & landmarks outdoors \\
blockfield & landmarks outdoors \\
cape & landmarks outdoors \\
cliff & landmarks outdoors \\
crevasse & landmarks outdoors \\
dune & landmarks outdoors \\
earth bank & landmarks outdoors \\
fell & landmarks outdoors \\
fumarole & landmarks outdoors \\
geyser & landmarks outdoors \\
glacier & landmarks outdoors \\
grassland & landmarks outdoors \\
heath & landmarks outdoors \\
isthmus & landmarks outdoors \\
moor & landmarks outdoors \\
mud & landmarks outdoors \\
peninsula & landmarks outdoors \\
reef & landmarks outdoors \\
rock & landmarks outdoors \\
sand & landmarks outdoors \\
shingle & landmarks outdoors \\
shoal & landmarks outdoors \\
sinkhole & landmarks outdoors \\
spring & landmarks outdoors \\
stone & landmarks outdoors \\
strait & landmarks outdoors \\
tundra & landmarks outdoors \\
valley & landmarks outdoors \\
water & landmarks outdoors \\
wetland & landmarks outdoors \\
\bottomrule
\end{tabular}
\end{table}

\begin{table}[!t]
\caption{Mappings for OSM Depth 1 = place}\label{taba:27}
\centering
\begin{tabular}{l|l}
\toprule
\textbf{OSM tag} & \textbf{FS manual match} \\
\midrule
continent & landmarks outdoors \\
ocean & landmarks outdoors \\
polder & landmarks outdoors \\
sea & landmarks outdoors \\
\bottomrule
\end{tabular}
\end{table}

\begin{table}[!t]
\caption{Mappings for OSM Depth 1 = power}\label{taba:28}
\centering
\begin{tabular}{l|l}
\toprule
\textbf{OSM tag} & \textbf{FS manual match} \\
\midrule
catenary mast & travel transportation \\
\bottomrule
\end{tabular}
\end{table}

\begin{table}[!t]
\caption{Mappings for OSM Depth 1 = railway}\label{taba:29}
\centering
\begin{tabular}{l|l}
\toprule
\textbf{OSM tag} & \textbf{FS manual match} \\
\midrule
funicular & travel transportation \\
light rail & travel transportation \\
roundhouse & travel transportation \\
switch & travel transportation \\
\bottomrule
\end{tabular}
\end{table}

\begin{table}[!t]
\caption{Mappings for OSM Depth 1 = route}\label{taba:30}
\centering
\begin{tabular}{l|l}
\toprule
\textbf{OSM tag} & \textbf{FS manual match} \\
\midrule
foot & travel transportation \\
subway & travel transportation \\
tracks & travel transportation \\
tram & travel transportation \\
\bottomrule
\end{tabular}
\end{table}

\begin{table}[!t]
\caption{Mappings for OSM Depth 1 = shop}\label{taba:31}
\centering
\begin{tabular}{l|l}
\toprule
\textbf{OSM tag} & \textbf{FS manual match} \\
\midrule
boat & retail \\
brewing supplies & retail \\
country store & retail \\
energy & retail \\
hairdresser supply & retail \\
military surplus & retail \\
outpost & retail \\
religion & retail \\
trade & retail \\
vacant & retail \\
\bottomrule
\end{tabular}
\end{table}

\begin{table}[!t]
\caption{Mappings for OSM Depth 1 = tourism}\label{taba:32}
\centering
\begin{tabular}{l|l}
\toprule
\textbf{OSM tag} & \textbf{FS manual match} \\
\midrule
attraction & arts entertainment \\
\bottomrule
\end{tabular}
\end{table}

\begin{table}[!t]
\caption{Mappings for OSM Depth 1 = water}\label{taba:33}
\centering
\begin{tabular}{l|l}
\toprule
\textbf{OSM tag} & \textbf{FS manual match} \\
\midrule
ditch & landmarks outdoors \\
fish pass & landmarks outdoors \\
lagoon & landmarks outdoors \\
moat & landmarks outdoors \\
\bottomrule
\end{tabular}
\end{table}

\begin{table}[!t]
\caption{Mappings for OSM Depth 1 = waterway}\label{taba:34}
\centering
\begin{tabular}{l|l}
\toprule
\textbf{OSM tag} & \textbf{FS manual match} \\
\midrule
fairway & landmarks outdoors \\
pressurised & landmarks outdoors \\
tidal channel & landmarks outdoors \\
water point & landmarks outdoors \\
\bottomrule
\end{tabular}
\end{table}

\newpage
\section{Prompt Design for OSM–FS Mapping}\label{asec:prompt}

\begin{lstlisting}[caption={Prompt construction for mapping OSM to FS categories (no fallback - no example).}, label={lst:prompt2}]
def ask_gpt_to_choose_prompt_2(osm_tag, osm_path, osm_desc, candidates, k):
    
    prompt = f"""
    I want to map an OpenStreetMap (OSM) point of interest (POI) to the most appropriate FourSquare (FS) POI (tag).

    Rules:
    1. If an FS tag exactly matches the OSM tag (same name or clear synonym), choose that FS tag directly.
    2. If no exact match exists, choose the FS tag that is the most specific and precise category in which the OSM tag could reasonably be classified.
    - Do not just pick the most semantically similar.
    - Prefer the FS tag that fully contains the concept of the OSM tag, even if its wording is broader.
    3. Always answer with exactly one FS tag name from the provided list, nothing else.

    OSM Tag: {osm_tag}  
    OSM Tag Description: {osm_desc}  
    OSM Tag Categorisation in OSM: {osm_path}  

    Here are the {k} most relevant FS tags (with their descriptions):  
    {candidates.to_string(index=False)}  

    Question: Which FS tag best matches the OSM tag?  
    Answer only with the FS tag name. """ 

    response = client.chat.completions.create(
        model="gpt-4o-mini",  
        messages=[{"role": "user", "content": prompt}],
        temperature=0
    )

    return response.choices[0].message.content.strip()
\end{lstlisting}

\begin{lstlisting}[caption={Prompt construction with fallback categories for mapping OSM to FS categories (fallback - no example).}, label={lst:prompt3}]
def ask_gpt_to_choose_prompt_3(osm_tag, osm_path, osm_desc, candidates, k):
    
    prompt = f"""
    I want to map an OpenStreetMap (OSM) point of interest (POI) to the most appropriate FourSquare (FS) POI (tag).

    Rules:
    1. If an FS tag exactly matches the OSM tag (same name or clear synonym), choose that FS tag directly.
    2. If no exact match exists, choose the FS tag that is the most specific and precise category in which the OSM tag could reasonably be classified.
    - Do not just pick the most semantically similar.
    - Prefer the FS tag that fully contains the concept of the OSM tag, even if its wording is broader.
    3. If none of the provided FS tags are a good fit, you may instead choose one category from the following broader fallback categories:
    - landmarks outdoors
    - business professional services
    - travel transportation
    - community government
    - retail
    - sports recreation
    - health medicine
    - arts entertainment
    - dining drinking
    - event
    4. Always answer with exactly one FS tag name from the provided list or, if necessary, one fallback category, nothing else.

    OSM Tag: {osm_tag}  
    OSM Tag Description: {osm_desc}  
    OSM Tag Categorisation in OSM: {osm_path}  

    Here are the {k} most relevant FS tags (with their descriptions):  
    {candidates.to_string(index=False)}  

    Question: Which FS tag best matches the OSM tag?  
    Answer only with the FS tag name. """ 

    response = client.chat.completions.create(
        model="gpt-4o-mini",  
        messages=[{"role": "user", "content": prompt}],
        temperature=0
    )

    return response.choices[0].message.content.strip()
\end{lstlisting}
\begin{lstlisting}[caption={Prompt construction with worked example for mapping OSM to FS categories (no fallback - example).}, label={lst:prompt5}]
def ask_gpt_to_choose_prompt_5(osm_tag, osm_path, osm_desc, candidates, k):
    prompt = f"""
    I want to map an OpenStreetMap (OSM) point of interest (POI) to the most appropriate FourSquare (FS) POI (tag).

    Rules:
    1. If an FS tag exactly matches the OSM tag (same name or clear synonym), choose that FS tag directly.
    2. If no exact match exists, select the FS tag that is the most precise parent category that fully contains the OSM tag.
       - Do not just pick the most semantically similar wording.
       - Prefer a broader FS tag that logically includes the OSM tag concept, even if it is less specific.
       - Exclude FS tags that are related but do not actually contain the OSM concept.
    3. Think step by step: first check for an exact match, then find the correct parent category.
    4. Output only one FS tag name from the provided list. No explanations, no extra text.

    Example:
    OSM Tag: "sea"  
    OSM Tag Description: "A large body of salt water part of, or connected to, an ocean."  
    OSM Tag Categorisation in OSM: "place > sea"  

    FS Candidates:  
    ['lake', 'bay', 'bathing area', 'dive spot', 'island', 'surf spot', 'waterfront', 'river', 'landmarks outdoors', 'boat launch']

    Correct Answer: landmarks outdoors

    Now, process the following case:

    OSM Tag: {osm_tag}  
    OSM Tag Description: {osm_desc}  
    OSM Tag Categorisation in OSM: {osm_path}  

    Here are the {k} most relevant FS tags (with their descriptions):  
    {candidates.to_string(index=False)}  

    Question: Which FS tag best matches the OSM tag?  
    Answer only with the FS tag name.
    """

    response = client.chat.completions.create(
        model="gpt-4o-mini",  
        messages=[{"role": "user", "content": prompt}],
        temperature=0
    )

    return response.choices[0].message.content.strip()
\end{lstlisting}
\begin{lstlisting}[caption={Prompt combining fallback categories and worked example for mapping OSM to FS categories (fallback - example).}, label={lst:prompt6}]
def ask_gpt_to_choose_prompt_6(osm_tag, osm_path, osm_desc, candidates, k):
    prompt = f"""
    I want to map an OpenStreetMap (OSM) point of interest (POI) to the most appropriate FourSquare (FS) POI (tag).

    Rules:
    1. If an FS tag exactly matches the OSM tag (same name or clear synonym), choose that FS tag directly.
    2. If no exact match exists, select the FS tag that is the most precise parent category that fully contains the OSM tag.
       - Do not just pick the most semantically similar wording.
       - Prefer a broader FS tag that logically includes the OSM tag concept.
       - Exclude FS tags that are related but do not actually contain the OSM concept.
    3. If none of the {k} FS candidates are suitable, then choose from the following broader FS categories:
       - landmarks outdoors
       - business professional services
       - travel transportation
       - community government
       - retail
       - sports recreation
       - health medicine
       - arts entertainment
       - dining drinking
       - event
    4. Think step by step: first check for an exact match, then find the correct parent category, and only if needed, fall back to the broader FS categories above.
    5. Output only one FS tag name from the provided list (either from the {k} candidates or from the broader categories).
       Do not add explanations, reasoning, or extra text.

    Example:
    OSM Tag: "sea"  
    OSM Tag Description: "A large body of salt water part of, or connected to, an ocean."  
    OSM Tag Categorisation in OSM: "place > sea"  

    FS Candidates:  
    ['lake', 'bay', 'bathing area', 'dive spot', 'island', 'surf spot', 'waterfront', 'river', 'landmarks outdoors', 'boat launch']

    Correct Answer: landmarks outdoors 

    Now, process the following case:

    OSM Tag: {osm_tag}  
    OSM Tag Description: {osm_desc}  
    OSM Tag Categorisation in OSM: {osm_path}  

    Here are the {k} most relevant FS tags (with their descriptions):  
    {candidates.to_string(index=False)}  

    Question: Which FS tag best matches the OSM tag?  
    Answer only with the FS tag name.
    """

    response = client.chat.completions.create(
        model="gpt-4o-mini",  
        messages=[{"role": "user", "content": prompt}],
        temperature=0
    )

    return response.choices[0].message.content.strip()
\end{lstlisting}

\end{document}